\documentclass[review]{elsarticle}
\usepackage[T1]{fontenc}
\usepackage[latin9]{inputenc}
\usepackage{graphicx}
\usepackage{caption}
\usepackage{subcaption}
\usepackage{fancybox}
\usepackage{fullpage}
\usepackage{hyperref}
\usepackage{color}
\usepackage{url,textcomp}
\usepackage{amsmath}
\usepackage{amsfonts}
\usepackage{amsthm}
\usepackage{amsbsy}
\usepackage{amssymb}
\usepackage[english]{babel}
\usepackage{enumerate}
\usepackage[section]{placeins}
\usepackage{lineno}
\modulolinenumbers[5]

\journal{Journal of  }

\begin{document}

\begin{frontmatter}

\title{Empirical Wavelet-based Estimation for Non-linear Additive Regression Models.}

\author{German A. Schnaidt Grez\footnote{Georgia Institute of Technology. Email:\url{gschnaidt@gatech.edu}}}
\address{755 Ferst Dr. NW, Atlanta, Georgia 30332}
\author{Brani Vidakovic\footnote{Georgia Institute of Technology. Email:\url{brani@gatech.edu}} }
\address{755 Ferst Dr. NW, Atlanta, Georgia 30332}

\begin{abstract}
Additive regression models are actively researched in the statistical field because of their usefulness in the analysis of responses determined by non-linear relationships with multivariate predictors. In this kind of statistical models, the response depends linearly on unknown functions of predictor variables and typically, the goal of the analysis is to make inference about these functions.

\medskip

In this paper, we consider the problem of Additive Regression with random designs from a novel viewpoint: we propose an estimator based on an orthogonal projection onto a multiresolution space using empirical wavelet coefficients that are fully data driven. In this setting, we derive a mean-square consistent estimator based on periodic wavelets on the interval $[0,1]$. For construction of the estimator, we assume that the joint distribution of predictors is non-zero and bounded on its support; We also assume that the functions belong to a Sobolev space and integrate to zero over the [0,1] interval, which guarantees model identifiability and convergence of the proposed method. Moreover, we provide the $\mathbb{L}_{2}$ risk analysis of the estimator and derive its convergence rate.

\medskip

Theoretically, we show that this approach achieves good convergence rates when the dimensionality of the problem is relatively low and the set of unknown functions is sufficiently smooth. In this approach, the results are obtained without the assumption of an equispaced design, a condition that is typically assumed in most wavelet-based procedures.

\medskip

Finally, we show practical results obtained from simulated data, demonstrating the potential applicability of our method in the problem of additive regression models with random designs.

\end{abstract}


\begin{keyword}
Wavelets\sep non-parametric regression\sep functional data analysis \sep robust statistical modeling
\end{keyword}

\end{frontmatter}

\linenumbers

\newpage

\section{Introduction}\label{Intro}

Additive regression models are popular in the statistical field because of their usefulness in the analysis of responses determined by non-linear relationships involving multivariate predictors. In this kind of statistical models, the response depends linearly on unknown functions of the predictors and typically, the goal of the analysis is to make inferences about these functions. This model has been extensively studied through the application of piecewise polynomial approximations, splines, marginal integration, as well as back-fitting or functional principal components. Chapter 15 of \cite{Ramsay2005}, Chapter 22 of \cite{Gyorfi2002} and \cite{Mammen2003}, \cite{Buja1989} and \cite{Hastie1990} feature thorough discussions of the issues related to fitting such models and provide a comprehensive overview and analysis of various estimation techniques for this problem.

\medskip

In general, the additive regression model relates a univariate response $Y$ to predictor variables $\textbf{X}\in \mathbb{R}^{p}\,,\,p\geq1$, via a set of unknown non-linear functions $\left\{f_{l}\,|\,f_{l}:\mathbb{R}\rightarrow R\,,\,l=1,...,p\right\}$. The functions $f_{l}$ may be assumed to have a specified parametric form (e.g. polynomial) or may be specified non-parametrically, simply as "smooth functions" that satisfy a set of constraints (e.g. belong to a certain functional space such as a Besov or Sobolev, Lipschitz continuity, spaces of functions with bounded derivatives, etc.). Though the parametric estimates may seem more attractive from the modeling perspective, they can have a major drawback: a parametric model automatically restricts the space of functions that is used to approximate the unknown regression function, regardless of the available data. As a result, when the elicited parametric family is not "close" to the assumed functional form the results obtained through the parametric approach can be misleading. For this reason, the non-parametric approach has gained more popularity in statistical research, providing a more general, flexible and robust approach in tasks of functional inference.

\medskip

In this paper we propose a linear functional estimator based on an orthogonal projection onto a specified multiresolution space $V_{J}$ using empirical wavelet coefficients that are fully data driven. Here, $V_{J}$ stands for the space spanned by the set of scaling functions of the form $\left\{\phi_{Jk}^{per},\,0\leq k \leq 2^{J}-1\right\}$, generated by a specified wavelet filter. Since we assume predictors $\textbf{X}\in \mathbb{R}^{p}\,,\,p\geq1$ are random with an unknown distribution, we introduce a kernel density estimator in the model to estimate its density. In this setting, we propose a mean-square consistent estimator for the constant term and the wavelet coefficients in the orthogonal series representation of the model. Our results are based on wavelets periodic on the interval $[0,1]$ and are derived under a set of assumptions that guarantee identifiability and convergence of the proposed estimator. Moreover, we derive convergence rates for the $\mathbb{L}_{2}$ risk and propose a practical choice for the multiresolution index $J$ to be used in the wavelet expansion. In this approach, we obtain stated results without the assumption of an equispaced design, a condition that is typically assumed in most wavelet-based procedures.

\medskip
Our choice of wavelets as an orthonormal basis is motivated by the fact that wavelets are well localized in both time and scale (frequency), and possess superb approximation properties for signals with rapid local changes such as discontinuities, cusps, sharp spikes, etc.. Moreover, the representation of these signals in the form of wavelet decompositions can be accurately done using only a few wavelet coefficients, enabling sparsity and dimensionality reduction. This adaptivity does not, in general, hold for other standard orthonormal bases (e.g. Fourier basis) which may require many compensating coefficients to describe signal discontinuities or local bursts. 
\medskip

We also illustrate practical results for the proposed estimator using different exemplary functions and random designs, under different sample sizes, demonstrating the suitability of the proposed methodology.

\medskip

As it was mentioned, additive regression models have been studied by many authors using a wide variety of approaches. The approaches include marginal integration, back-fitting, least squares (including penalized least squares), orthogonal series approximations, and local polynomials. Short descriptions of the most commonly used techniques are provided next:

\begin{enumerate}[(i)]
\item {\textbf{Marginal Integration}.} This method was proposed by Tjostheim and Auestad (1994)\cite{Tjostheim1994} and Linton and Nielsen (1995)\cite{Nielsen1995} and later generalized by Chen et al. (1996)\cite{Chen1996}. The marginal integration idea is based on the estimation of the effects of each function in the model using sample averages of kernel functions by keeping a variable of interest fixed at each observed sample point, while changing the remaining ones. This method has been shown to produce good results in simulation studies (Sperlich et al., 1999)\cite{Sperlich1999}. However, the marginal integration performance over finite samples tends to be inadequate when the dimension of the predictors is large. In particular, the bias-variance trade-off of the estimator in this case is challenging: for a given bandwidth there may be too few data points $\textbf{x}_{i}$ for any given $\textbf{x}$, which inflates the estimator variance and reduces its numerical stability. On the other hand, choosing larger bandwidth may reduce the variability but also enlarge the bias.

\item {\textbf{Back-fitting}.} This approach was first introduced by Buja et al. (1989)\cite{Buja1989a} and further developed by Hastie and Tibshirani (1990)\cite{Hastie1990a}. This technique uses nonparametric regression to estimate each additive component, and then updates the preliminary estimates. This process continues in an iterative fashion until convergence. One of the drawbacks of this method is that it has been proven to be theoretically challenging to analize. In this context, Opsomer and Ruppert (1997)\cite{Opsomer1997} investigated the properties of a version of back-fitting, and found that the estimator was not oracle efficient\footnote{An oracle efficient estimator is such that each component of the model can be estimated with the same convergence rate as if the rest of the model components were known.}. Later on, Mammen et al. (1999)\cite{Mammen1999} and Mammen and Park (2006)\cite{Mammen2006} proposed ways to modify the backfitting approach to produce estimators with better statistical properties such as oracle efficiency and asymptotic normality, and also free of the curse of dimensionality. Even though this is a popular method, it has been shown that its efficiency decreases when the unknown functions are observed at nonequispaced locations.

\item {\textbf{Series based methods using wavelets}.} One important benefit of wavelets is that they are able to adapt to unknown smoothness of functions (Donoho et al. (1995)\cite{Donoho1995}). Most of the work using wavelets is based on the requirement of equally spaced measurements (e.g. at equal time intervals or a certain response observed on a regularly spaced grid). Antoniadis et al. (1997)\cite{Antoniadis1997a} propose a method using interpolations and averaging; based on the observed sample, the function is approximated at equally spaced dyadic points. In this context, most of the methods that use this kind of approach lead to wavelet coefficients that can be computed via a matrix transformation of the original data and are formulated in terms of a continuous wavelet transformation applied to a constant piecewise interpolation of the observed samples. Pensky and Vidakovic (2001)\cite{Pensky2001} propose a method that uses a probabilistic model on the design of the independent variables and can be applied to non-equally spaced designs (NESD). Their approach is based on a linear wavelet-based estimator that is similar to the wavelet modification of the Nadaraja-Watson estimator (Antoniadis et al. (1994)). In the same context, Amato and Antoniadis (2001)\cite{Amato2001} propose a wavelet series estimator based on tensor wavelet series and a regularization rule that guarantees an adaptive solution to the estimation problem in the presence of NESD.

\item {\textbf{Other methods based on wavelets}.} Different approaches from the previously described that are wavelet-based have been also investigated. Donoho et al. (1992)\cite{Donoho1992} proposed an estimator that is the solution of a penalized Least squares optimization problem preventing the problem of ill-conditioned design matrices. Zhang and Wong (2003) proposed a two-stage wavelet thresholding procedure using local polynomial fitting and marginal integration for the estimation of the additive components. Their method is adaptive to different degrees of smoothness of the components and has good asymptotic properties. Later on Sardy and Tseng (2004)\cite{Sardy2004} proposed a non-linear smoother and non-linear back-fitting algorithm that is based on \verb"WaveShrink", modeling each function in the model as a parsimonious expansion on a wavelet basis that is further subjected to variable selection (i.e. which wavelets to use in the expansion) via non-linear shrinkage.
\end{enumerate}

As was discussed before in the context of the application of wavelets to the problem of additive models in NESD, another possibility is just simply ignore the nonequispaced condition on the predictors and apply the wavelet methods directly to the observed sample. Even though this might seem a somewhat crude approach, we will show that it is possible to implement this procedure via a relatively simple algorithm, obtaining good statistical properties and estimation results.

\subsection{About Periodic Wavelets}\label{wavelets}

For the implementation of the functional estimator, we choose periodic wavelets as an orthonormal basis. Even though this kind of wavelets exhibit poor behaviour near the boundaries (when the analyzed function is not periodic, high amplitude wavelet coefficients are generated in the neighborhood of the boundaries) they are typically used due to the relatively simple numerical implementation and compact support. Also, as was suggested by Johnstone (1994), this simplification affects only a small number of wavelet coefficients at each resolution level.

\medskip

Periodic wavelets in $[0,1]$ are defined by a modification of the standard scaling and wavelet functions:
\begin{eqnarray}
 & \phi^{per}_{j,k}(x)=\sum_{l \in \mathbb{Z}}\phi_{j,k}(x-l) \,,\\
 & \psi^{per}_{j,k}(x)=\sum_{l \in \mathbb{Z}}\psi_{j,k}(x-l)\,.
\end{eqnarray}

It is possible to show, as in \cite{Restrepo1996}, that $\left\{ \phi^{per}_{j,k}(x), 0\leq k \leq 2^{j}-1 , j\geq 0\right\}$ constitutes an orthonormal basis
for $\mathbb{L}_{2}[0,1]$. Consequently, $\cup_{j=0}^{\infty} V_{j}^{per}=\mathbb{L}_{2}[0,1]$, where $V_{j}^{per}$ is the space spanned by $\left\{ \phi^{per}_{j,k}(x), 0\leq k \leq 2^{j}-1 \right\}$. This allows to represent a function $f$ with support in $[0,1]$ as:

\begin{equation}\label{eq:1b}
f(x)=\langle f(x),\phi^{per}_{0,0}(x) \rangle \phi^{per}_{0,0}(x) + \sum_{j\geq 0}\sum_{k=0}^{2^{j}-1}\langle f(x),\psi^{per}_{j,k}(x) \rangle \psi^{per}_{j,k}(x)\,.
\end{equation}

\medskip
Also, for a fixed $j=J$, we can obtain an orthogonal projection of $f(x)$ onto $V_{J}$ denoted as $\textbf{P}_{J}(f(x))$ given by:

\begin{equation}\label{eq:1c}
\textbf{P}_{J}(f(x))=\sum_{k=0}^{2^{J}-1}\langle f(x),\phi^{per}_{J,k}(x) \rangle \phi^{per}_{J,k}(x)
\end{equation}
Since periodized wavelets provide a basis for $\mathbb{L}^{2}([0,1])$, we have that $\parallel f(x) - \textbf{P}_{J}(f(x)) \parallel_{2} \rightarrow 0$ as $ J \rightarrow  \infty $. Also, it can be shown that $\parallel f(x) - \textbf{P}_{J}(f(x)) \parallel_{\infty} \rightarrow 0$ as $ J \rightarrow  \infty $. Therefore, we can see that $\textbf{P}_{J}(f(x))$ uniformly converges to $f$ as $J \rightarrow \infty$.
\medskip
Similarly, as discussed in \cite{Daubechies1992} it is possible to assess the approximation error for a certain density of interest $f$ using a truncated projection (i.e. for a certain chosen detail space $J$). For example, using the $s$-th Sobolev norm of a function defined as:

\begin{equation}
\parallel f(x) \parallel_{H^{s}}=\sqrt{\int(1+|x|^{2})^{s}|f(x)|^{2}dx}\,,
\end{equation}

one defines the $H^{s}$ sobolev space, as the space that consists of all functions $f$ whose s-Sobolev norm exists and is finite. As it is shown in \cite{Daubechies1992}:

\begin{equation}\label{eq:1d}
\parallel f(x) - \textbf{P}_{J}(f(x)) \parallel_{2} \leq 2^{-J\cdot s}\cdot \parallel f \parallel _{H^{s}[0,1]}\,.
\end{equation}

From (\ref{eq:1d}), for a pre-specified $\epsilon>0$ one can choose $J$ such that $\parallel f(x) - \textbf{P}_{J}(f(x)) \parallel_{2} \leq \epsilon$. In fact, a possible choice of
J could be:

\begin{equation}\label{eq:1e}
J \geq -\lceil \frac{1}{s} \log_{2}(\frac{\epsilon}{\parallel f \parallel _{H^{s}[0,1]}}) \rceil\,.
\end{equation}

Therefore, it is possible to approximate a desired function to arbitrary precision using the MRA generated by a wavelet basis.


%


\newpage
\section{Wavelet-based Estimation in Additive Regression Models}\label{ARM}

Suppose that instead of the typical linear regression model $y=\sum_{j=1}^{p}\beta_{j}x_{j}+\beta_{0}+\epsilon$ which assumes linearity
in the predictors $\textbf{x}=\left(x_{1},...,x_{p}\right)$, we have the following:
\begin{eqnarray}
\nonumber
f(\textbf{x})&=&\beta_{0}+f_{A}(\textbf{x})+\sigma\cdot\epsilon \\
 &=&\beta_{0}+\sum_{j=1}^{p}f_{j}(x_{j})+\sigma\cdot\epsilon \label{eq:3.1}
\end{eqnarray}
where $\epsilon$, independent of $\textbf{x}$, $\mathbb{E}[\epsilon]=0$, $\mathbb{E}[\epsilon^{2}]=1$, $\sigma>0$, $\sigma<\infty$. Similarly, $\textbf{x}_{i} \mathop{\sim}\limits^{\text{iid}} h(\textbf{x})$, an unknown design density of observations and $\left\{f_{1}(),...,f_{p}()\right\}$ are unknown functions to be estimated.

\medskip

\subsection{Problem statement and derivation of the Estimator}\label{ARMStatementDerivation}

Suppose that we are able to observe a sample $\{y_{i}=f(\textbf{x}_{i}),\textbf{x}_{i} \}_{i=1}^{n}$ where $\textbf{x}_{1},...,\textbf{x}_{n} \mathop{\sim}\limits^{iid} h(\textbf{x})$. We are interested in
estimating $\beta_{0}$ and $\{f_{1}(),...,f_{p}()\}$. For simplicity (without loss of generality) and identifiability, we assume:
\begin{enumerate}[]
\item{(A1)}\label{A1} The density $h(\textbf{x})$ is of the continuous type and has support in $[0,1]^{p}$. Also, we assume $\exists \epsilon_{h}>0\,\,$ s.t. $\,h(\textbf{x})\geq \epsilon_{h}$ $\, \forall \textbf{x} \in [0,1]^{p}$.
\item{(A2)}\label{A2} For $k=1,...,p$, $\int_{0}^{1}f_{k}(x)dx_{k}=0$.
\item{(A3)}\label{A3} For $k=1,...,p$, $\mathop{sup}\limits_{x\in [0,1]}|f_{k}(x)|\leq M_{k}< \infty$ and $\mathop{inf}\limits_{x\in [0,1]}\left\{f_{k}(x)\right\}\geq m_{k}> -\infty$. This implies that for $k=1,...,p$, $f_{k}\in \mathbb{L}_{2}([0,1])$.
\item{(A4)}\label{A4} The design density $h()$ belongs to a generalized Holder class of functions of the form:
\begin{equation}
\mathbb{H}(\beta,L)=\{h : |\partial^{\boldsymbol{\alpha}}h(\textbf{x})-\partial^{\boldsymbol{\alpha}}h(\textbf{y})|\leq L\parallel \textbf{x}-\textbf{y}\parallel^{\beta - |\boldsymbol{\alpha}|}_{1} ,\, \forall \boldsymbol{\alpha}\in\mathbb{N}^{p}, \,s.t.\, |\boldsymbol{\alpha}|=\lfloor \beta \rfloor, \,\forall \,\textbf{x},\textbf{y} \in [0,1]^{p}\} \label{eq:3.2}
\end{equation}
where $\partial^{\boldsymbol{\alpha}}f := \partial_{1}^{\alpha_{1}}\cdot...\cdot\partial_{p}^{\alpha_{p}}f =
\frac{\partial^{|\boldsymbol{\alpha}|}f}{\partial x_{1}^{\alpha_{1}}\cdot...\cdot\partial x_{p}^{\alpha_{p}}}$, and $|\boldsymbol{\alpha}|:=\sum_{j=1}^{p}\alpha_{j}$. Also, suppose that $|\partial^{\boldsymbol{\alpha}}h|\leq M_{h}$, for all $\textbf{x}\in [0,1]^{p}$ and $|\boldsymbol{\alpha}|\leq \lfloor \beta \rfloor$.

\medskip

\item{(A5)}\label{A6} The density $h(\textbf{x})$ is uniformly bounded in $[0,1]^{p}$, that is, $\forall \textbf{x}\in [0,1]^{p}, \,  |h(\textbf{x})|\leq M$, $M<\infty$.
\end{enumerate}
\medskip

\medskip

Furthermore, since $\left\{ \phi^{per}_{j,k}(x), k \in [0,2^{j}-1] , j\geq 0\right\}$ spans $\mathbb{L}_{2}([0,1])$, each of the functions in \ref{eq:3.1} can be represented as:
\begin{equation}\label{eq:3.3}
f_{l}(x)=\sum_{j\geq 0}\sum_{k=0}^{2^{j}-1}c_{jk}^{(l)}\cdot\phi_{jk}^{per}(x),\quad l=1,...,p\,,
\end{equation}
where $c_{jk}^{(l)}$ denotes the $j,k-$th wavelet coefficient of the $l-$th function in the model. Similarly, for some fixed $J$ that $f_{l,J}(x),\, l=1,...,p$ is the orthogonal projection of $f_{l}(x)$, onto the multiresolution space. Therefore, $f_{l,J}(x)$ can be expressed as:
\begin{equation}\label{eq:3.4}
f_{l,J}(x)=\sum_{k=0}^{2^{J}-1}c_{Jk}^{(l)}\cdot\phi_{Jk}^{per}(x),\quad l=1,...,p\,,
\end{equation}
where:

\begin{equation}\label{eq:3.5}
c_{Jk}^{(l)}=\langle f_{l}(x),\phi_{Jk}^{per}(x)\rangle=\int_{0}^{1}f_{l}(x)\phi_{Jk}^{per}(x)dx,\quad l=1,...,p\,.
\end{equation}
Based on the model (\ref{eq:3.1}) and (\ref{eq:3.4}), it is possible to approximate $f(\textbf{x})$ by an orthogonal projection $f_{J}(\textbf{x})$ onto the multiresolution space
spanned by the set of scaling functions $\left\{ \phi^{per}_{J,k}(x), 0\leq k \leq 2^{J}-1\right\}$, by approximating each of the functions $f_{l}()$ as described above. Therefore, $f_{J}(\textbf{x})$ can be expressed as:

\begin{equation}\label{eq:3.6}
f_{J}(\textbf{x})=\beta_{0}+\sum_{l=1}^{p}\sum_{k=0}^{2^{J}-1}c_{Jk}^{(l)}\phi_{Jk}^{per}(x)
\end{equation}
Now, the goal is for a pre-specified multiresolution index $J$, to use the observed samples to estimate the unknown constant $\beta_{0}$ and the orthogonal projections of the functions $f_{l,J}(x),\, l=1,...,p$.

\subsubsection*{Remarks}
\begin{enumerate}[(i)]
\item \label{A7} Note that the scaling function $\phi(x)$ for the wavelet basis $\left\{ \phi^{per}_{j,k}(x), k \in [0,2^{j}-1] , j\geq 0\right\}$ is absolutely integrable in $\mathbb{R}$. Therefore, $\int_{\mathbb{R}}|\phi(x)|dx=C_{\phi}<\infty$.
\item \label{A7} Also, from the above conditions, the variance of the response $y(\textbf{x})$ is bounded for every $\textbf{x}\in \mathbb{R}^{p}$.
\item The assumption that the support of the random vector $\textbf{X}$ is $[0,1]^{p}$ can be always satisfied by carrying out appropriate monotone increasing transformations of each dimensional component, even in the case when the support before transformation is unbounded. In practice, it would be sufficient to transform the empirical support to $[0,1]^{p}$.
\end{enumerate}

\subsubsection{Derivation of the estimator for $\beta_{0}$}

From the model definition presented in (\ref{eq:3.1}), and assumption (\textbf{A2}) we have that:

\begin{eqnarray}
\nonumber
\int_{[0,1]^{p}} (\beta_{0}+\sum_{l=1}^{p}f_{l}(x_{l}))d\textbf{x} &=& \beta_{0}+\sum_{l=1}^{p}\int_{0}^{1}f_{l}(x_{l})dx_{l} \\
 &=& \beta_{0} \label{eq:3.7}
\end{eqnarray}
Therefore, under assumptions (\textbf{A1}) and the last result, it is possible to obtain $\beta_{0}$ as:

\begin{equation}\label{eq:3.8}
\beta_{0} = \mathbb{E}_{\textbf{X},\epsilon}\left[\frac{f(\textbf{X})}{h(\textbf{X})}\right]\,.
\end{equation}
Indeed,

\begin{eqnarray}
\nonumber
\mathbb{E}_{\textbf{X},\epsilon}\left[\frac{f(\textbf{X})}{h(\textbf{X})}\right] &=&  \mathbb{E}_{\textbf{X},\epsilon}\left[\frac{\beta_{0}+\sum_{j=1}^{p}f_{j}(X_{j})+\sigma\cdot\epsilon}{h(\textbf{X})} \right] \\
\nonumber
&=&  \beta_{0} + \mathbb{E}_{\textbf{X}}\left[\frac{\sum_{j=1}^{p}f_{j}(X_{j})}{h(\textbf{X})} \right]\\
\nonumber
&=&  \beta_{0}\,.
\end{eqnarray}
As a result of (\ref{eq:3.8}), a natural data-driven estimator of $\beta_{0}$ is

\begin{equation}\label{eq:3.9}
\hat{\beta}_{0}=\frac{1}{n}\sum_{i=1}^{n}\frac{y_{i}}{\hat{h}_{n}(\textbf{x}_{i})}\,,
\end{equation}
where $\hat{h}_{n}()$ is a suitable non-parametric density estimator of $h()$, e.g. a kernel density estimator.

\subsubsection{Derivation of the estimator for the wavelet coefficients $c_{Jk}^{(l)}$}

Based on the multiresolution space spanned by the orthonormal functions $\left\{\phi^{per}_{J,k}(x)\right\}$,
(\ref{eq:3.5}) and assumption (\textbf{A2}), the wavelet coefficients for each functional can be represented as:

\begin{equation}\label{eq:3.10}
c_{Jk}^{(l)} = \int_{0}^{1}f_{l}(x_{l})\phi_{Jk}^{per}(x_{l})dx_{l}\,.
\end{equation}
Expanding the right-hand-side (rhs) of the last equation, we get:

\begin{eqnarray}
\nonumber
 \int_{0}^{1}f_{l}(x_{l})\phi_{Jk}^{per}(x_{l})dx_{l} &=& \int_{0}^{1}f_{l}(x_{l})\left(\phi_{Jk}^{per}(x_{l})-2^{-\frac{J}{2}}\right)dx_{l} \\
\nonumber
  &=& \int_{0}^{1}\left(\beta_{0}+\sum_{j=1}^{p} f_{j}(x_{j})\right)\left(\phi_{Jk}^{per}(x_{l})-2^{-\frac{J}{2}}\right)dx_{l} \\
 \nonumber
  &=& \int_{[0,1]^{p-1}}\int_{0}^{1}\left(\beta_{0}+\sum_{j=1}^{p} f_{j}(x_{j})\right)\left(\phi_{Jk}^{per}(x_{l})-2^{-\frac{J}{2}}\right)dx_{l}d\textbf{x}_{(-l)}\,,
\end{eqnarray}
where $\textbf{x}_{(-l)}$ corresponds to the random vector $\textbf{x}$ without the $l-$th entry. It is easy to see that (\ref{eq:3.10}) holds because of assumption (\textbf{A2}) and the fact that $\int_{0}^{1}\phi_{Jk}^{per}(x)dx=2^{-\frac{J}{2}}$. The proof for this last claim can be found in \ref{proof:periodic}.

\medskip

Now, if we consider (\textbf{A1}), we can see that an alternative way to express (\ref{eq:3.10}) could be:

\begin{equation}\label{eq:3.11}
c_{Jk}^{(l)} = \mathbb{E}_{\textbf{X},\epsilon}\left[\frac{f(\textbf{X})(\phi_{Jk}^{per}(x_{l})-2^{-\frac{J}{2}})}{h(\textbf{X})}\right]\,.
\end{equation}
Indeed,

\begin{eqnarray}
\nonumber
\mathbb{E}_{\textbf{X},\epsilon}\left[\frac{f(\textbf{X})(\phi_{Jk}^{per}(x_{l})-2^{-\frac{J}{2}})}{h(\textbf{X})}\right] &=& \mathbb{E}_{\textbf{X},\epsilon}\left[\frac{(\beta_{0}+\sum_{j=1}^{p}f_{j}(X_{j})+\sigma\cdot\epsilon)(\phi_{Jk}^{per}(x_{l})-2^{-\frac{J}{2}})}{h(\textbf{X})}\right] \\
\nonumber
&=& \int_{[0,1]^{p}}\left(\beta_{0}+\sum_{j=1}^{p} f_{j}(x_{j})\right)\left(\phi_{Jk}^{per}(x_{l})-2^{-\frac{J}{2}}\right)d\textbf{x} \\
\nonumber
&=& c_{Jk}^{(l)}\,.
\end{eqnarray}
From (\ref{eq:3.11}), similarly as for $\beta_{0}$, we obtain a natural data-driven estimator of $c_{Jk}^{(l)}$ as:

\begin{equation}\label{eq:3.12}\,.
\hat{c}_{Jk}^{(l)}=\frac{1}{n}\sum_{i=1}^{n}\frac{y_{i}\left(\phi_{Jk}^{per}(x_{il})-2^{-\frac{J}{2}} \right)}{\hat{h}_{n}(\textbf{x}_{i})}
\end{equation}

\subsection{Asymptotic Properties of the Estimator}\label{ARMProperties}

In this section, we study the asymptotic properties of the estimates proposed in (\ref{eq:3.9}) and (\ref{eq:3.12}) and propose necessary and sufficient conditions for
the pointwise mean squared consistency of the estimator, under assumptions (\textbf{A1})-(\textbf{A5}).

\subsubsection{Unbiasedness and Consistency of $\hat{\beta}_{0}$}\label{BetahatConsistency}

Next, we analyze the asymptotic behavior of the estimator $\hat{\beta}_{0}$ assuming assumptions (\textbf{Ak1})-(\textbf{Ak4}) stated in \ref{proof:kernelconsistency} hold.

\medskip

\subsubsection*{Asymptotic Behavior of $\mathbb{E}(\hat{\beta}_{0})$}

From (\ref{eq:4.2}) and the hierarchy of convergence for random variables, it follows that for a fixed $\textbf{x}$, $\hat{h}_{n}(\textbf{x})\mathop{\rightarrow}\limits^{\mathbb{D}}h(\textbf{x})$. Let's consider now a function $g:[\epsilon_{h},M]\rightarrow[0,B_{h}]$, for $\epsilon_{h}>0$, $B_{h}<\infty$,  defined as $g(\hat{h}_{n}(\textbf{x}))=\frac{1}{\hat{h}_{n}(\textbf{x})}$. Since $\hat{h}_{n}(\textbf{x})$ satisfies (\textbf{A5})-(\textbf{A6}), $g(h)$ is bounded and continuous, which implies:

\begin{equation}\label{eq:4.3}
\mathbb{E}\left[ \frac{1}{\hat{h}_{n}(\textbf{x})}\right] \mathop{\rightarrow}\limits_{n\rightarrow\infty} \frac{1}{h(\textbf{x})}\,.
\end{equation}
In fact, since $g(\hat{h}_{n}(\textbf{x}))=\frac{1}{\hat{h}_{n}(\textbf{x})}$ is continuous in $(0,\infty)$ and admits infinitely many derivatives , by using a Taylor series expansion around $h(\textbf{x})$ and results (\ref{eq:k.4}) and (\ref{eq:k.5}), it is possible to obtain:

\begin{eqnarray}
\nonumber
\left|\mathbb{E}_{\textbf{X}_{1},...,\textbf{X}_{n}}\left[\left(\frac{1}{\hat{h}_{n}(\textbf{x})}\right)^{k}-\left(\frac{1}{h(\textbf{x})}\right)^{k}\right]\right| &\leq& \frac{1}{\epsilon_{h}^{k+2}}\left\{|Bias(\hat{h}_{n}(\textbf{x}))|+Var(\hat{h}_{n}(\textbf{x}))+
Bias(\hat{h}_{n}(\textbf{x}))^{2} \right\} \\
&\leq& C\left\{\delta^{\beta}+\frac{1}{n\delta^{p}}+\delta^{2\beta} \right\} \label{eq:4.3b}\,,
\end{eqnarray}
for $k\geq1$ and a sufficiently large $C>0$ (independent of $n$, $\delta$).

\medskip

Therefore, under the choice $\delta \sim n^{-\frac{1}{2\beta+p}}$, $\mathbb{E}\left[\left(\frac{1}{\hat{h}_{n}(\textbf{x})}\right)^{k}\right]$ converges to $\left(\frac{1}{h(\textbf{x})}\right)^{k}$ at a rate $\sim n^{-\frac{\beta}{2\beta+p}}$ for $k\geq1$. Here the expectation is taken with respect to the joint density of the iid sample.

\medskip

Similarly, the last result leads to:

\begin{equation}\label{eq:4.3c}
\mathbb{E}_{\textbf{X}_{1},...,\textbf{X}_{n}}\left[\left(\frac{1}{\hat{h}_{n}(\textbf{x})}-\frac{1}{h(\textbf{x})}\right)^{2}\right]\longrightarrow 0\,,
\end{equation}
as $ n \rightarrow \infty$ at a rate $\sim n^{-\frac{2\beta}{2\beta+p}}$.

\medskip

Now, letting $\textbf{x}$ to be random, using conditional expectation it is possible to obtain:

\begin{equation}\label{eq:4.4}
\mathbb{E}\left[\frac{1}{\hat{h}_{n}(\textbf{X})}\right]=\mathbb{E}_{\textbf{X}}\left[\mathbb{E}_{\textbf{X}_{1},...,\textbf{X}_{n}|\textbf{X}}\left(\frac{1}{\hat{h}_{n}(\textbf{x})}|\textbf{X} \right) \right]\,.
\end{equation}
From (\ref{eq:4.3}) and the last result, the dominated convergence theorem implies:

\begin{equation}\label{eq:4.5}
\mathbb{E}\left[ \frac{1}{\hat{h}_{n}(\textbf{x})}\right] \mathop{\longrightarrow}\limits_{n\rightarrow\infty} 1
\end{equation}
Using the definition of $\hat{\beta}_{0}$ and the model (\ref{eq:3.1}), we obtain:

\begin{eqnarray}
\nonumber
\mathbb{E}\left[\hat{\beta}_{0} \right] &=& \beta_{0}+\mathbb{E}\left[\frac{\sum_{l=1}^{p}f_{l}(X_{l})}{\hat{h}_{n}(\textbf{X})}\right] \\
\nonumber
 &=& \beta_{0}+\mathbb{E}_{\textbf{X}}\left[\mathbb{E}_{\textbf{X}_{1},...,\textbf{X}_{n}|\textbf{X}} \left(\frac{\sum_{l=1}^{p}f_{l}(X_{l})}{\hat{h}_{n}(\textbf{X})}|\textbf{X}\right)\right] \\ &=& \beta_{0}+\mathbb{E}_{\textbf{X}}\left[\sum_{l=1}^{p}f_{l}(X_{l})\cdot\mathbb{E}_{\textbf{X}_{1},...,\textbf{X}_{n}|\textbf{X}} \left(\frac{1}{\hat{h}_{n}(\textbf{X})}|\textbf{X}\right) \right]\,. \label{eq:4.6}
\end{eqnarray}
Therefore, from (\ref{eq:4.3})-(\ref{eq:4.5}) and under (\textbf{A2}),(\textbf{A3}), the dominated convergence leads to:

\begin{equation} \label{eq:4.7}
\mathbb{E}\left[\hat{\beta}_{0}\right]\mathop{\longrightarrow}\limits_{n\rightarrow\infty} \beta_{0}\,,
\end{equation}
which shows that $\hat{\beta}_{0}$ is asymptotically unbiased for $\beta_{0}$.

\subsubsection*{Asymptotic Behavior of Var($\hat{\beta}_{0}$)}

From the definition of $\hat{\beta}_{0}$ and (\ref{eq:3.1}), we can see that:

\begin{eqnarray}
\nonumber
Var(\hat{\beta}_{0}) &=& \frac{1}{n}Var\left(\frac{Y}{\hat{h}_{n}(\textbf{X})}\right) \\
\nonumber
& \leq & \frac{1}{n}\mathbb{E}\left[\frac{Y^{2}}{\hat{h}_{n}(\textbf{X})^{2}}\right] \\
& \leq & \frac{1}{n}\mathbb{E}_{\textbf{X}}\left[\mathbb{E}_{\textbf{X}_{1},...,\textbf{X}_{n}|\textbf{X}=\textbf{x}} \left(\frac{Y^{2}}{\hat{h}_{n}(\textbf{X})^{2}} | \textbf{X}=\textbf{x}\right)\right]\,. \label{eq:4.8}
\end{eqnarray}
Now, if $n\rightarrow\infty$, from conditions (\textbf{A2}) and (\textbf{A3}), and the dominated convergence theorem, it follows:

\begin{equation} \label{eq:4.9}
\mathbb{E}_{\textbf{X}}\left[\mathbb{E}_{\textbf{X}_{1},...,\textbf{X}_{n}|\textbf{X}=\textbf{x}} \left(\frac{Y^{2}}{\hat{h}_{n}(\textbf{X})^{2}} | \textbf{X}=\textbf{x}\right)\right] \mathop{\rightarrow}\limits_{n\longrightarrow\infty} \mathbb{E}\left[\frac{Y^{2}}{h(\textbf{X})^{2}}\right]\,.
\end{equation}

Thus,

\begin{equation} \label{eq:4.10}
Var(\hat{\beta}_{0}) \mathop{\longrightarrow}\limits_{n\rightarrow\infty} 0\,,
\end{equation}
provided $ \mathbb{E}\left[\frac{Y^{2}}{h(\textbf{X})^{2}}\right]<\infty$.

\medskip

Finally, putting together (\ref{eq:4.7}) and (\ref{eq:4.10}) we obtain that $\hat{\beta}_{0}$ is consistent for $\beta_{0}$.

\subsubsection{Unbiasedness and Consistency of the $\hat{c}_{Jk}^{(l)}$}\label{CJKConsistency}

In this section, we study the asymptotic behavior of the wavelet coefficient estimators $\hat{c}_{Jk}^{(l)}$ for a fixed $J$, assuming that conditions (\textbf{A1})-(\textbf{A5}) and (\textbf{Ak1})-(\textbf{Ak4}) hold.

\subsubsection*{Asymptotic Behavior of $\mathbb{E}(\hat{c}_{Jk}^{(l)})$}

For a fixed $J$, $l=1,...,p$, and $k=0,...,2^{J}-1$,we have that $\hat{c}_{Jk}^{(l)}=\frac{1}{n}\sum_{i=1}^{n}\frac{y_{i}\left(\phi_{Jk}^{per}(x_{il})-2^{-\frac{J}{2}} \right)}{\hat{h}_{n}(\textbf{x}_{i})}$. Therefore,

\begin{equation}\label{eq:4.11}
\mathbb{E}\left[\hat{c}_{Jk}^{(l)} \right]= \mathbb{E}\left[\frac{Y\phi_{Jk}^{per}(X_{l})}{\hat{h}_{n}(\textbf{X})}\right]-2^{-\frac{J}{2}}\mathbb{E}\left[\hat{\beta}_{0}\right]\,.
\end{equation}
Following the same argument as in the case of the asymptotic behavior of $\hat{\beta}_{0}$, we find that the first term of (\ref{eq:4.11}) can be represented as:

\begin{eqnarray}
\nonumber
\mathbb{E}\left[\frac{Y\phi_{Jk}^{per}(X_{l})}{\hat{h}_{n}(\textbf{X})}\right] &=& \mathbb{E}_{\textbf{X}}\left[\mathbb{E}_{\textbf{X}_{1},...,\textbf{X}_{n}|\textbf{X}}\left(\frac{Y\phi_{Jk}^{per}(X_{l})}{\hat{h}_{n}(\textbf{X})}| \textbf{X}\right)\right] \\
\nonumber
&=& \mathbb{E}_{\textbf{X}}\left[Y\phi_{Jk}^{per}(X_{l})\cdot\mathbb{E}_{\textbf{X}_{1},...,\textbf{X}_{n}|\textbf{X}}\left(\frac{1}{\hat{h}_{n}(\textbf{X})}| \textbf{X}\right)\right]\,. \label{eq:4.12}
\end{eqnarray}
Since $J$ is assumed fixed and (\textbf{A3}) holds, by the dominated convergence theorem, it follows that:

\begin{equation}\label{eq:4.13}
\mathbb{E}_{\textbf{X}}\left[Y\phi_{Jk}^{per}(X_{l})\cdot\mathbb{E}_{\textbf{X}_{1},...,\textbf{X}_{n}|\textbf{X}}\left(\frac{1}{\hat{h}_{n}(\textbf{X})}| \textbf{X}\right)\right] \mathop{\longrightarrow}\limits_{n\rightarrow\infty} \mathbb{E}\left[\frac{Y\phi_{Jk}^{per}(X_{l})}{h(\textbf{X})}\right]\,.
\end{equation}
Furthermore, by (\textbf{A3}) and (\ref{eq:proof1}):

\begin{eqnarray}
\nonumber
\mathbb{E}\left[\frac{Y\phi_{Jk}^{per}(X_{l})}{h(\textbf{X})}\right] &=& \int_{[0,1]^{p}}\left(\beta_{0}+\sum_{j=1}^{p}f_{j}(x_{j})\right)\phi_{Jk}^{per}(x_{l})d\textbf{x} \\
\nonumber
&=& \int_{0}^{1}f_{l}(x_{l})\phi_{Jk}^{per}(x_{l})dx_{l}+2^{-\frac{J}{2}}\beta_{0} \\
&=& c_{Jk}^{(l)}+2^{-\frac{J}{2}}\beta_{0}\,. \label{eq:4.14}
\end{eqnarray}
Finally, putting together the last result and (\ref{eq:4.7}), it follows:

\begin{equation}\label{eq:4.15}
\mathbb{E}\left[\hat{c}_{Jk}^{(l)}\right] \mathop{\longrightarrow}\limits_{n\rightarrow\infty} c_{Jk}^{(l)}\,,
\end{equation}
which shows that the wavelet coefficient estimators $\hat{c}_{Jk}^{(l)}$ are asymptotically unbiased, for $J$ fixed, $l=1,...,p$, and $k=0,...,2^{J}-1$.

\subsubsection*{Asymptotic Behavior of Var($\hat{c}_{Jk}^{(l)}$)}

For a fixed $J$, $l=1,...,p$ and $k=0,...,2^{J}-1$,  $\hat{c}_{Jk}^{(l)}=\frac{1}{n}\sum_{i=1}^{n}\frac{Y_{i}\left(\phi_{Jk}^{per}(X_{il})-2^{-\frac{J}{2}} \right)}{\hat{h}_{n}(\textbf{x}_{i})}$, the variance of $\hat{c}_{Jk}^{(l)}$ is given by:

\begin{eqnarray}
\nonumber 
Var\left(\hat{c}_{Jk}^{(l)} \right) &=& Var\left(\frac{1}{n}\sum_{i=1}^{n}\frac{Y_{i}\phi_{Jk}^{per}(X_{il})}{\hat{h}_{n}(\textbf{X}_{i})}-2^{-\frac{J}{2}}\hat{\beta}_{0}\right) \\
 &=& \frac{1}{n}Var\left(\frac{Y\phi_{Jk}^{per}(X_{l})}{\hat{h}_{n}(\textbf{X})}\right)+2^{-J}Var\left(\hat{\beta}_{0}\right)-2Cov\left(\frac{1}{n}
 \sum_{i=1}^{n}\frac{Y_{i}\phi_{Jk}^{per}(X_{il})}{\hat{h}_{n}(\textbf{X}_{i})}\,,\,2^{-\frac{J}{2}}\hat{\beta}_{0}\right) \label{eq:4.16} \\
 \nonumber
 &=& \frac{1}{n}V_{c1}+2^{-J}V_{c2}+2V_{c3}\,.
\end{eqnarray}
By using the model defined in (\ref{eq:3.1}) we find that for $V_{c1}=\frac{1}{n}Var\left(\frac{Y\phi_{Jk}^{per}(X_{l})}{\hat{h}_{n}(\textbf{X})}\right)$:

\begin{equation}
\resizebox{.99 \textwidth}{!}{$
\nonumber
V_{c1} = Var_{\textbf{X}}\left(\mathbb{E}_{\textbf{X}_{1},...,\textbf{X}_{n}|\textbf{X}}
\left[\frac{Y\phi_{Jk}^{per}(X_{l})}{\hat{h}_{n}(\textbf{X})}|\textbf{X} \right]\right)+\mathbb{E}_{\textbf{X}}\left[Var_{\textbf{X}_{1},...,\textbf{X}_{n}|\textbf{X}}\left(\frac{Y\phi_{Jk}^{per}(X_{l})}{\hat{h}_{n}(\textbf{X})}|\textbf{X} \right) \right]$}\,,
\end{equation}
\begin{equation} \label{eq:4.17}
\resizebox{.9 \textwidth}{!}{$
  = Var_{\textbf{X}}\left(Y\phi_{Jk}^{per}(X_{l})\cdot\mathbb{E}_{\textbf{X}_{1},...,\textbf{X}_{n}|\textbf{X}}
\left[\frac{1}{\hat{h}_{n}(\textbf{X})}|\textbf{X} \right]\right)+\mathbb{E}_{\textbf{X}}\left[(Y\phi_{Jk}^{per}(X_{l}))^{2}\cdot Var_{\textbf{X}_{1},...,\textbf{X}_{n}|\textbf{X}}\left(\frac{1}{\hat{h}_{n}(\textbf{X})}|\textbf{X} \right) \right]
$}\,.
\end{equation}
By the dominated convergence theorem, it follows:

\begin{equation}\label{eq:4.18}
Var\left(\frac{Y\phi_{Jk}^{per}(X_{l})}{\hat{h}_{n}(\textbf{X})}\right)  \mathop{\longrightarrow}\limits_{n\rightarrow\infty} Var\left(\frac{Y\phi_{Jk}^{per}(X_{l})}{h(\textbf{X})}\right)
\end{equation}
where the last result holds since:

\begin{eqnarray}
\nonumber 
\mathbb{E}_{\textbf{X}_{1},...,\textbf{X}_{n}|\textbf{X}=\textbf{x}}
\left[\frac{1}{\hat{h}_{n}(\textbf{X})}|\textbf{X}=\textbf{x} \right] &\mathop{\longrightarrow}\limits_{n\rightarrow\infty}& \frac{1}{h(\textbf{x})}\,,\,\text{and}  \\
\nonumber
Var_{\textbf{X}_{1},...,\textbf{X}_{n}|\textbf{X}=\textbf{x}}\left(\frac{1}{\hat{h}_{n}(\textbf{X})}|\textbf{X}=\textbf{x} \right) &\mathop{\longrightarrow}\limits_{n\rightarrow\infty}& 0\,.
\end{eqnarray}
This implies,

\begin{equation}\label{eq:4.18b}
\frac{1}{n}Var\left(\frac{Y\phi_{Jk}^{per}(X_{l})}{\hat{h}_{n}(\textbf{X})}\right) \mathop{\longrightarrow}\limits_{n\rightarrow\infty} 0\,.
\end{equation}
\subsubsection*{Proposition 2} \label{prop2}

Let us suppose that conditions (\textbf{A1})-(\textbf{A5}) and (\textbf{Ak1})-(\textbf{Ak4}) hold hold. Then:

\begin{equation}\label{eq:4.19}
\mathbb{E}\left[\left(\frac{Y\phi_{Jk}^{per}(X_{l})}{h(\textbf{X})} \right)^{2} \right]\leq C(\beta_{0},p,\sigma^{2},M_{f})\cdot\left\{
 \frac{1}{\epsilon_{h}}\left(\lceil\log_{2}(\frac{1}{\epsilon_{h}})\rceil-1 \right)+\frac{1}{\lceil \log_{2}(\frac{1}{\epsilon_{h}})\rceil}\right\}\,,
\end{equation}
where $C(\beta_{0},p,\sigma^{2},M_{f})=\left(p\cdot M_{f}+|\beta_{0}|\right)^{2}+\sigma^{2}$. This result shows that $Var\left(\frac{Y\phi_{Jk}^{per}(X_{l})}{h(\textbf{X})}\right)$ is bounded from above, provided $p<\infty$, $\sigma^{2}<\infty$ and conditions (\textbf{A1})-(\textbf{A5}) and (\textbf{Ak1})-(\textbf{Ak4}) hold. Therefore,
\begin{equation}
\nonumber
Var\left(\frac{Y\phi_{Jk}^{per}(X_{l})}{\hat{h}_{n}(\textbf{X})}\right)  \mathop{\rightarrow}\limits_{n\rightarrow\infty} Var\left(\frac{Y\phi_{Jk}^{per}(X_{l})}{h(\textbf{X})}\right)<\infty\,.
\end{equation}
\medskip
The proof can be found in \ref{proof:ExpUpper}.

\medskip

Similarly, as for $V_{c1}$, let's consider the behavior of $V_{c3}=Cov\left(\frac{1}{n}\sum_{i=1}^{n}\frac{Y_{i}\phi_{Jk}^{per}(X_{il})}{\hat{h}_{n}(\textbf{X}_{i})}\,,\,2^{-\frac{J}{2}}\hat{\beta}_{0}\right)$. Using the covariance definition and the iid assumption for the sample $\{y_{i}=f(\textbf{x}_{i}),\textbf{x}_{i} \}_{i=1}^{n}$, it follows that:

\begin{equation}\label{eq:4.20}
  V_{c3} = \frac{2^{-\frac{J}{2}}}{n^{2}}\left\{\sum_{i=1}^{n}Cov\left(\frac{Y_{i}\phi_{Jk}^{per}(X_{il})}{\hat{h}_{n}(\textbf{X}_{i})}\,,\,\frac{Y_{i}}{\hat{h}_{n}(\textbf{X}_{i})} \right)+
  \mathop{\sum_{i=1}^{n}\sum_{j=1}^{n}}\limits_{i\neq j}Cov\left(\frac{Y_{i}\phi_{Jk}^{per}(X_{il})}{\hat{h}_{n}(\textbf{X}_{i})}\,,\,\frac{Y_{j}}{\hat{h}_{n}(\textbf{X}_{j})} \right)\right\}\,.
\end{equation}
\subsubsection*{Proposition 3} \label{prop3}

Let us suppose assumptions (\textbf{A1})-(\textbf{A5}) and (\textbf{Ak1})-(\textbf{Ak4}) are satisfied. The following results hold:

\begin{equation}\label{eq:4.21}
Cov\left(\frac{1}{n}\sum_{i=1}^{n}\frac{Y_{i}\phi_{Jk}^{per}(X_{il})}{\hat{h}_{n}(\textbf{X}_{i})}\,,\,2^{-\frac{J}{2}}\hat{\beta}_{0}\right) \mathop{\longrightarrow}\limits_{n\rightarrow\infty} 0\,,
\end{equation}
which further implies that for any fixed $J$, $l=1,...,p$, and $k=0,...,2^{J}-1$,

\begin{equation}\label{eq:4.21b}
Cov\left(\hat{\beta}_{0}\,,\,\hat{c}_{Jk}^{(l)}\right)\mathop{\longrightarrow}\limits_{n\rightarrow\infty} 0\,.
\end{equation}
The corresponding proofs can be found in \ref{proof:asympUncorrelation}.

\medskip
\medskip
\medskip

Putting together (\ref{eq:4.10}), (\ref{eq:4.18b}) and (\ref{eq:4.21}) it follows that for a fixed $J$, $l=1,...,p$, and $k=0,...,2^{J}-1$:

\begin{equation}\label{eq:4.22}
Var\left(\hat{c}_{Jk}^{(l)}\right) \mathop{\longrightarrow}\limits_{n\rightarrow\infty} 0\,.
\end{equation}
Finally, from (\ref{eq:4.15}) and (\ref{eq:4.22}) we get that for a fixed $J$, $l=1,...,p$, and $k=0,...,2^{J}-1$, $\hat{c}_{Jk}^{(l)}$ is consistent for $c_{Jk}^{(l)}$.

\subsubsection{Unbiasedness and Consistency of $\hat{f}_{J}(\textbf{x})$}\label{fJConsistency}

From (\ref{eq:3.6}), we have that $f_{J}(\textbf{x})=\beta_{0}+\sum_{l=1}^{p}\sum_{k=0}^{2^{J}-1}c_{Jk}^{(l)}\phi_{Jk}^{per}(x_{l})$. If results (\ref{eq:3.9}) and (\ref{eq:3.12}) are substituted in the expression for $f_{J}(\textbf{x})$, the data-driven estimator can be expressed as:

\begin{equation}
\nonumber
\hat{f}_{J}(\textbf{x})=\hat{\beta}_{0}+\sum_{l=1}^{p}\sum_{k=0}^{2^{J}-1}\hat{c}_{Jk}^{(l)}\phi_{Jk}^{per}(x_{l})\,.
\end{equation}
Since both $\hat{\beta}_{0}$ and $\hat{c}_{Jk}^{(l)}$ are asymptotically unbiased, it follows:

\begin{equation}\label{eq:4.23}
\mathbb{E}\left[\hat{f}_{J}(\textbf{x})\right] \mathop{\rightarrow}\limits_{n\rightarrow\infty}  f_{J}(\textbf{x})\,,\,\text{and}
\end{equation}
\begin{equation}
Var\left(\hat{f}_{J}(\textbf{x})\right)=Var\left(\hat{\beta}_{0}\right) + Var\left(\sum_{l=1}^{p}\sum_{k=0}^{2^{J}-1}\hat{c}_{Jk}^{(l)}\phi_{Jk}^{per}(x_{l})\right) +2Cov\left(\hat{\beta}_{0}\,,\,\sum_{l=1}^{p}\sum_{k=0}^{2^{J}-1}\hat{c}_{Jk}^{(l)}\phi_{Jk}^{per}(x_{l})\right)\,.\label{eq:4.23b}
\end{equation}
In order to show that $Var\left(\hat{f}_{J}(\textbf{x})\right) \mathop{\rightarrow}\limits_{n\rightarrow\infty} 0$, we just need to prove that the second term of the expression (\ref{eq:4.23b}) goes to zero as $n \rightarrow \infty$. This can be seen from (\ref{eq:4.10}) and (\ref{eq:4.21b}).

\medskip

\subsubsection*{Proposition 4} \label{prop4}

For any $s\neq k$, $s,k=0,...,2^{J}-1$ and fixed $J$, under the stated assumptions:

\begin{equation}\label{eq:4.24}
Cov\left(\hat{c}_{Jk}^{(l)}\,,\,\hat{c}_{Js}^{(l)} \right) \mathop{\longrightarrow}\limits_{n\rightarrow\infty} 0\,.
\end{equation}
The proof can be found in \ref{proof:asympUncorrelationCoeff}.

\medskip

From  (\ref{eq:4.24}) it follows:

\begin{equation}\label{eq:4.25}
Var\left(\sum_{l=1}^{p}\sum_{k=0}^{2^{J}-1}\hat{c}_{Jk}^{(l)}\phi_{Jk}^{per}(x_{l})\right)\mathop{\longrightarrow}\limits_{n\rightarrow\infty} 0\,.
\end{equation}
\medskip

Finally, from (\ref{eq:4.10}), (\ref{eq:4.21b}) and (\ref{eq:4.25}), it is clear that $Var\left(\hat{f}_{J}(\textbf{x})\right) \mathop{\rightarrow}\limits_{n\rightarrow\infty} 0$. This result together with (\ref{eq:4.23}) implies that:

\begin{equation}\label{eq:4.26}
\hat{f}_{J}(\textbf{x}) \mathop{\longrightarrow}\limits^{\mathbb{P}} f_{J}(\textbf{x})\,.
\end{equation}
Therefore, the estimator $\hat{f}_{J}(\textbf{x})$ is consistent for $f_{J}(\textbf{x})$.

\subsubsection*{Remarks}

\begin{enumerate}[(i)]
  \item The results and derivations presented in Propositions \textbf{1}-\textbf{4}, indicate that our estimator $\hat{f}_{J}(\textbf{x})$ suffers from the course of dimensionality. In fact, the dependence from the dimension $p$ of the random covariates \textbf{x} influence in both the convergence rate of the density estimator $\hat{h}_{n}(\textbf{x})$ and the constant $C(\beta_{0},p,\sigma^{2},M_{f})$.
  \item As can be seen from this section results, one of the key assumptions used to show consistency of the estimates $\hat{f}_{J}(\textbf{x})$, $\hat{c}_{Jk}^{(l)}$ and $\hat{\beta}_{0}$, is that the multiresolution index $J$ is kept fixed. This ensures that $|\phi_{Jk}^{per}(x)|<\infty$, which enables the use of the dominated convergence theorem. Nonetheless, as it will be shown in the next section, it is possible to relax such assumption, enabling that $J=J(n)$ and furthermore, $J(n)\rightarrow\infty$ as $n\rightarrow\infty$.
\end{enumerate}

\subsection{$\mathbb{L}_{2}$ Risk Analysis of the Estimator $\hat{f}_{J}(\textbf{x})$}\label{RiskARM}

In the last section, we showed that the estimates $\hat{f}_{J}(\textbf{x})$, $\hat{c}_{Jk}^{(l)}$ and $\hat{\beta}_{0}$ are unbiased and consistent for $f_{J}(\textbf{x})$, $c_{Jk}^{(l)}$ and $\beta_{0}$ respectively. In this section we provide a brief $\mathbb{L}_{2}$ risk analysis for the model estimate $f_{J}(\textbf{x})$ and we show that $R(\hat{f}_{J},f)=\mathbf{E}\left[||\hat{f}_{J}(\textbf{x})-f(\textbf{x})||^{2}_{2} \right]$ converges to zero as $n\rightarrow\infty$.

\medskip
As it will be demonstrated next, the rate of convergence of $\hat{f}_{J}(\textbf{x})$ is influenced by the convergence properties of the kernel density estimator $\hat{h}_{n}(\textbf{x})$ and the smoothness properties of the set $\left\{ \phi^{per}_{J,k}(x), 0\leq k \leq 2^{J}-1\right\}$ generated by the scaling function $\phi(x)$, together with the functions $\left\{f_{l}(x) \right\}_{l=1}^{p}$ that define the additive model.

\medskip

From the definition of $\hat{f}_{J}(\textbf{x})$ and Cauchy-Schwartz inequality, it follows:

\begin{equation}\label{eq:Risk1}
\mathbb{E}\left[||\hat{f}_{J}(\textbf{x})-f(\textbf{x}) ||_{2}^{2}\right] \leq 2\left(\mathbb{E}\left[||\hat{f}_{J}(\textbf{x})-\mathbb{E}[\hat{f}_{J}(\textbf{x})]||_{2}^{2}\right] +
||\mathbb{E}[\hat{f}_{J}(\textbf{x})]-f(\textbf{x}) ||_{2}^{2}\right)
\end{equation}
Note that the first term on the rhs of (\ref{eq:Risk1}) corresponds to the variance of the estimate $\hat{f}_{J}(\textbf{x})$, while the second represents the square of the $bias(\hat{f}_{J}(\textbf{x}))$.

\medskip

\subsubsection*{Proposition 5} \label{prop5}
Assume conditions (\textbf{A1})-(\textbf{A5}) and (\textbf{Ak1})-(\textbf{Ak4}) are satisfied. Then for $J=J(n)$ it follows:

\begin{equation}\label{eq:Risk2}
\mathbb{E}\left[||\hat{f}_{J}(\textbf{x})-\mathbb{E}[\hat{f}_{J}(\textbf{x})]||_{2}^{2}\right] = \mathcal{O}\left(2^{J(n)}n^{-1}\right)\,.
\end{equation}
The corresponding proof can be found in \ref{proof:Prop5}.

\subsubsection*{Proposition 6} \label{prop6}

In addition to conditions (\textbf{A1})-(\textbf{A5}) and (\textbf{Ak1})-(\textbf{Ak4}), assume conditions 1-7 described in \ref{proof:Prop6} hold. Then:

\medskip

\begin{equation}\label{eq:Risk3}
||\mathbb{E}[\hat{f}_{J}(\textbf{x})]-f(\textbf{x}) ||_{2}^{2} = \mathcal{O}\left(2^{2J(n)}n^{-\frac{2\beta}{2\beta+p}}+2^{-2J(n)(N+1)}+n^{-\frac{\beta}{2\beta+p}}2^{-J(n)(N+1)}\right)\,.
\end{equation}
The corresponding proof can be found in \ref{proof:Prop6}.

\subsubsection*{Proposition 7} \label{prop7}

Define $\mathcal{F}=\left\{f\,|\,f_{l}\,\in\,L_{2}([0,1]),\,f_{l}\,\in\,W_{2}^{N+1}([0,1]),\,-\infty< m_{l}\leq f_{l}\leq M_{l}<\infty\right\}$, where $f(\textbf{x})=\beta_{0}+\sum_{l=1}^{p}f_{l}(x_{l})$, and $W_{2}^{N+1}([0,1])$ represents the space of functions that are twice-differentiable with $f^{(k)}_{l}\in \mathbb{L}_{2}([0,1])$, $k=1,2$. Suppose assumptions for Propositions 5 and 6 hold, and  conditions (\textbf{A1})-(\textbf{A5}) and (\textbf{Ak1})-(\textbf{Ak4}) are satisfied. Then, it follows:

\begin{equation}\label{eq:Risk4}
\mathop{\sup}\limits_{f\in \mathcal{F}}\left(\mathbb{E}\left[||\hat{f}_{J}(\textbf{x})-f(\textbf{x}) ||_{2}^{2}\right] \right)\leq \tilde{C}n^{-\left(\frac{2\beta}{2\beta+p}\right)\left(\frac{N+1}{N+3}\right)}\,,
\end{equation}
provided (\ref{eq:Risk2}) and (\ref{eq:Risk3}), and $J=J(n)$ such that $2^{J(n)}\simeq n^{\frac{2\beta}{(2\beta+p)(N+3)}}$.

\medskip

Also, it is possible to show:

\begin{eqnarray}
\label{eq:Risk5}
\mathbb{E}\left[||\hat{f}_{J}(\textbf{x})-\mathbb{E}[\hat{f}_{J}(\textbf{x})]||_{2}^{2}\right] & = & \mathcal{O}\left(n^{-\left(\frac{N+2}{N+3}\right)}n^{-\left(\frac{p}{2\beta+p}\right)}\right) \,,\,\text{and}\\
\label{eq:Risk6}
||\mathbb{E}[\hat{f}_{J}(\textbf{x})]-f(\textbf{x}) ||_{2}^{2} & = & \mathcal{O}\left(n^{-\left(\frac{2\beta}{2\beta+p}\right)\left(\frac{N+1}{N+3}\right)}\right)\,.
\end{eqnarray}
\medskip
The corresponding proofs can be found in \ref{proof:Prop7}.

\medskip

\subsubsection*{Remarks and comments} \label{remarks7}

\begin{enumerate}[(i)]
\item The additional assumptions described in \ref{proof:Prop6} are needed to use the wavelet approximation results presented in chapters 8-9 (Corollary 8.2) of \cite{Hardle1998}.
\item As proposed in \cite{Hardle1998}, the simplest way to obtain the wavelet approximation property utilized in the derivation of (\ref{eq:Risk3}) is by selecting a bounded and compactly supported scaling function $\phi$ to generate $\left\{ \phi^{per}_{J,k}(x), 0\leq k \leq 2^{J}-1 \right\}$.
\item In the derivations for the convergence rate for the estimator $\hat{f}_{J}(\textbf{x})$, the smoothness assumptions for the unknown functions $f_{l}$ and the wavelet scaling function $\phi$ play a key role. In this sense, the index $N$ corresponds to the minimum smoothness index among the unknown functions $\left\{f_{1},...,f_{p} \right\}$ and the scaling function $\phi$.
\item From (\ref{eq:Risk5}) and (\ref{eq:Risk6}), it holds that the variance term of the estimator $\hat{f}_{J}(\textbf{x})$, for large dimensions $p$ is influenced primarily by the smoothness properties of the functional space that contains $\left\{f_{l}(x)\,,\,l=1,...,p\right\}$ and the wavelet basis $\left\{ \phi^{per}_{J,k}(x), k =0,...,2^{J}-1\right\}$. Also, for $n$ sufficiently large, the bias term dominates in the risk decomposition of $\hat{f}_{J}(\textbf{x})$.
\item As a result of the introduction of the density estimator $\hat{h}_{n}(\textbf{x})$ in the model, $\hat{f}_{J}(\textbf{x})$ suffers from the curse of dimensionality. In particular, it is interesting to note that this effect affects only the bias term, since as $p\rightarrow\infty$, $\mathbb{E}\left[||\hat{f}_{J}(\textbf{x})-\mathbb{E}[\hat{f}_{J}(\textbf{x})]||_{2}^{2}\right] \rightarrow \mathcal{O}(n^{-\frac{7}{4}})$, for $N\geq1$.
\item An alternative way to show the mean square consistency of the estimator $\hat{f}_{J}(\textbf{x})$ is via Stone's theorem (details can be found in Theorem 4.1 \cite{Gyorfi2002}), by assuming a model with no intercept (i.e. $\beta_{0}=0$), and expressing the estimator as:
        \begin{equation}
        \nonumber
        \hat{f}_{J}(\textbf{x})=\sum_{i=1}^{n}W_{n,i}(\textbf{x})\cdot y_{i}\,,
        \end{equation}
    where $W_{n,i}(\textbf{x})=\sum_{l=1}^{p}\sum_{k=0}^{2^{J}-1}\left( \frac{\phi_{Jk}^{per}(X_{il})-2^{-\frac{J}{2}}}{n \cdot \hat{h}_{n}(\textbf{X}_{i})} \right)\phi_{Jk}^{per}(x_{l}) $. Then, the estimator is mean-square consistent provided the following conditions are satisfied:
    \begin{enumerate}
    \item \label{assStone1} For any $n$, $\exists \, c \in \mathbb{R}$ such that for every non-negative measurable function $f$ satisfying $\mathbb{E}f(\textbf{X})<\infty$, $\mathbb{E}\left\{\sum_{i=1}^{n}\left|W_{n,i}(\textbf{x})f(\textbf{X}_{i})\right|\right\}\leq c\,\mathbb{E}f(\textbf{X})$.
    \item \label{assStone2} For all $n$, $\exists\,D\geq 1$ such that $\mathbb{P}\left\{\sum_{i=1}^{n}\left|W_{n,i}(\textbf{x})\right|\leq D \right\}=1$.
    \item \label{assStone3} For all $a>0$, $\mathop{\lim}\limits_{n \rightarrow \infty} \mathbb{E}\left\{\sum_{i=1}^{n}\left|W_{n,i}(\textbf{x})\right|\mathbf{1}_{\left\{||\textbf{X}_{i}-\textbf{x}||>a\right\}}\right\}=0$.
    \item \label{assStone4} $ \sum_{i=1}^{n}W_{n,i}(\textbf{x}) \mathop{\rightarrow}\limits_{n \rightarrow \infty}^{\mathbb{P}} 1$.
    \item \label{assStone5} $\mathop{\lim}\limits_{n \rightarrow \infty} \mathbb{E}\left\{\sum_{i=1}^{n}W_{n,i}(\textbf{x})^{2}\right\}=0$.
    \end{enumerate}
    \item Indeed, for the estimator $\hat{f}_{J}(\textbf{x})$ conditions (a)-(e) are satisfied, provided assumptions (\textbf{A1})-(\textbf{A5}) and (\textbf{Ak1})-(\textbf{Ak4}) hold, and for all $\textbf{x}\in [0,1]^{p}$, $\mathop{\lim}\limits_{n \rightarrow \infty} \left|1-\frac{h(\textbf{x})}{\hat{h}_{n}(\textbf{x})} \right|=0$.
\end{enumerate}

\subsection{Implementation illustration and considerations}

\subsubsection*{Implementation illustration}

In this section, we illustrate the application of the proposed method in a controlled experiment. For this purpose, we choose the following functions for the construction of model (\ref{eq:3.1}):
\begin{eqnarray}
\nonumber
f_{1}(x) & = & \frac{1}{\sqrt{2}\sin\left(2\pi\,x \right)} \\
\nonumber
f_{2}(x) & = & 1-4\,|x-\frac{1}{2}| \\
\nonumber
f_{3}(x) & = & -\cos\left(4\pi\,x +1\right) \\
\nonumber
f_{4}(x) & = & 8\,\left(x-\frac{1}{2} \right)^{2}-\frac{2}{3} \\
\nonumber
f_{5}(x) & = & \frac{1}{\sqrt{2}}\cos\left(2\pi\,x \right) \\
\nonumber
f_{6}(x) & = & \frac{1}{\sqrt{2}}\cos\left(4\pi\,x \right)
\end{eqnarray}
The estimator $\hat{f}_{J}(\textbf{x})$ was obtained using a box-type kernel with a bandwidth given by $\delta(n)=3n^{-\frac{1}{2+p}}$. For the multiresolution space index $J$, we chose $J(n)=4+\lfloor0.3\,\log_{2}(n)\rfloor$. The selection of the wavelet filter was Daubechies with 6 vanishing moments and the sample sizes used for this illustration were $n=512, \,4096\,\,\text{and}\,\, 8192$.

\medskip

Similarly, the noise in the model was defined to be gaussian with zero mean and variance given by $\sigma^{2}=0.45$. This led to a Signal to Noise Ratio (SNR) of approximately 8.6; Finally, the joint distribution for the predictors $\textbf{X}_{1},...,\textbf{X}_{n}$ was generated by independent $\mathcal{U}(0,1)$ and a $Beta(\frac{3}{2},\frac{3}{2})$ random variables along each dimension. For the evaluation of the scaling functions $\phi_{Jk}^{per}$ we used Daubechies-Lagarias's algorithm.

\begin{figure}[!htb]
   \centering
       \includegraphics[scale=0.8]{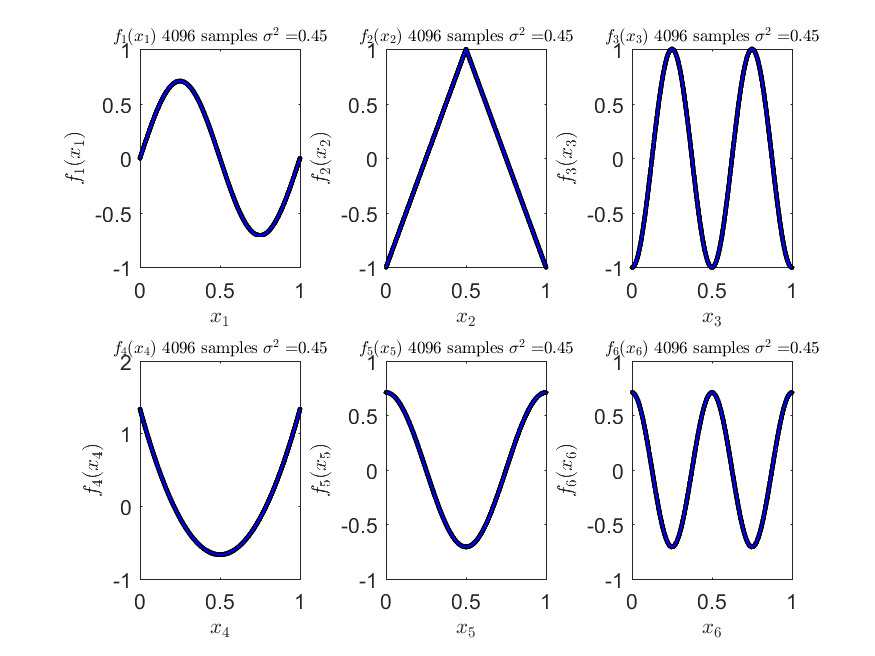}
       \caption{Additive functions used for illustration of the Estimator $\hat{f}_{J}(\textbf{x})$.}
       \label{fig:Functions4096UnifDesign_Efro}
\end{figure}

\begin{figure}[!htb]
   \centering
   \begin{subfigure}[b]{0.85\textwidth}
       \includegraphics[width=\textwidth]{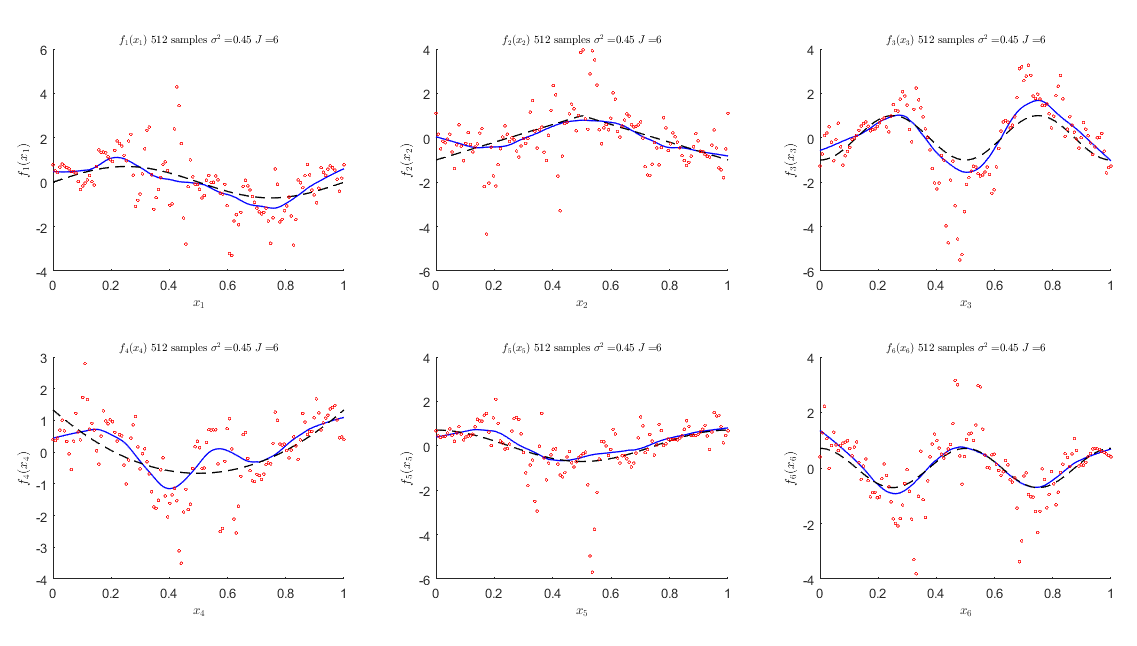}
       \caption{}
       \label{fig:FunctEst512BetaDesign_Efro}
   \end{subfigure}\vfill
   \begin{subfigure}[b]{0.85\textwidth}
       \includegraphics[width=\textwidth]{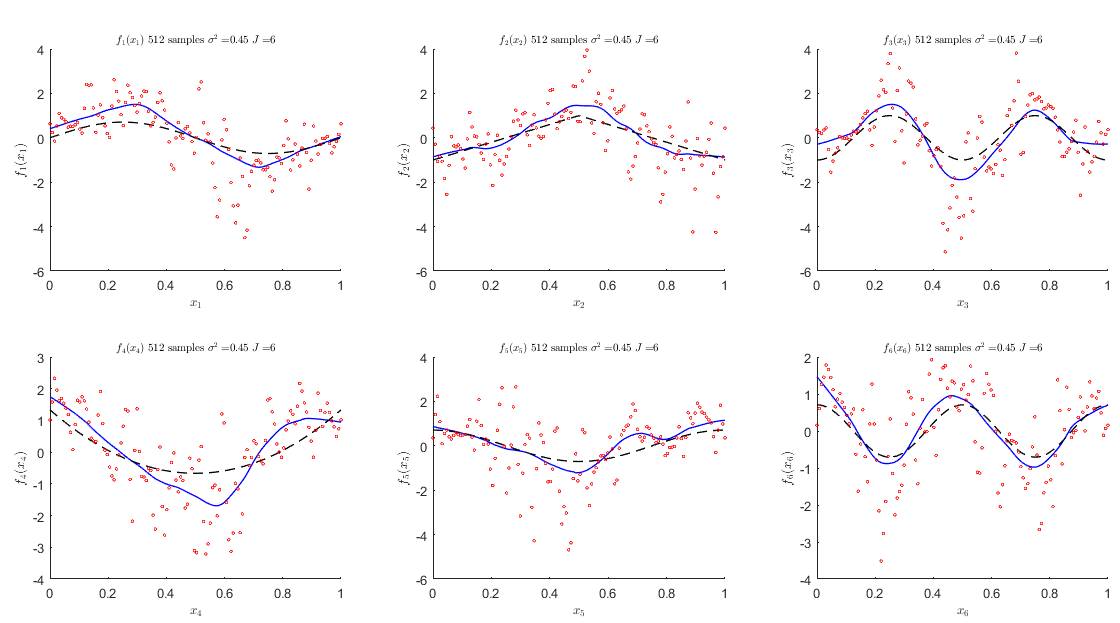}
       \caption{}
       \label{fig:FunctEst512UnifDesign_Efro}
    \end{subfigure}
\caption{Functions estimation for (a) $Beta(\frac{3}{2},\frac{3}{2})$and (b) $\mathcal{U}(0,1)$  designs, for $n=512$ samples. In red, the estimated
function values at each sample point; In black-dashed lines, the actual function shape; In blue lines, the smoothed version of the function values using lowess smoother.}
\label{fig:EstFuncsEfro512}
\end{figure}

\begin{figure}[!htb]
   \centering
   \begin{subfigure}[b]{0.85\textwidth}
       \includegraphics[width=\textwidth]{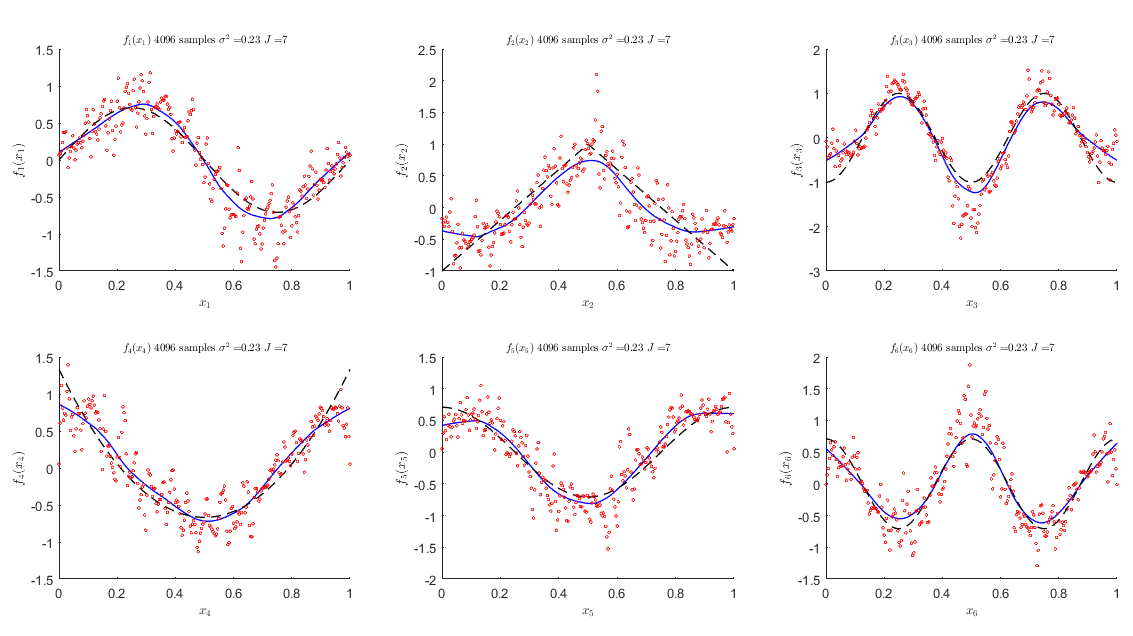}
       \caption{}
       \label{fig:FunctEst4096BetaDesign_Efro}
   \end{subfigure}\vfill
   \begin{subfigure}[b]{0.85\textwidth}
       \includegraphics[width=\textwidth]{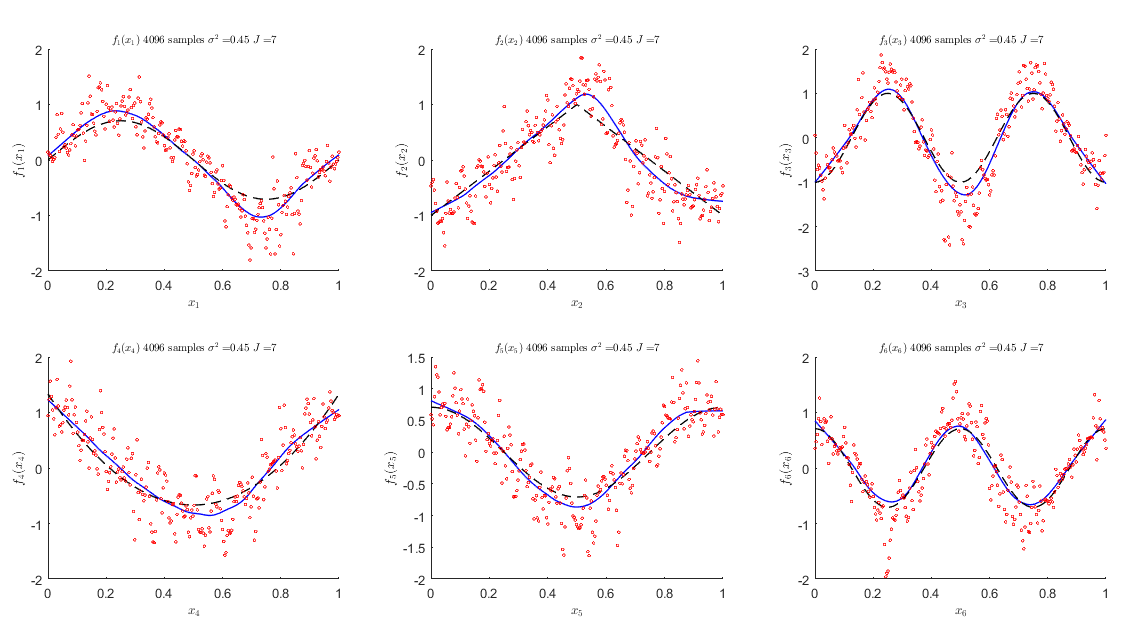}
       \caption{}
       \label{fig:FunctEst4096UnifDesign_Efro}
    \end{subfigure}
\caption{Functions estimation for (a) $Beta(\frac{3}{2},\frac{3}{2})$ and (b) $\mathcal{U}(0,1)$  designs, for $n=4096$ samples. In red, the estimated
function values at each sample point; In black-dashed lines, the actual function shape; In blue lines, the smoothed version of the function values using lowess smoother.}
\label{fig:EstFuncsEfro4096}
\end{figure}

\begin{figure}[!htb]
   \centering
   \begin{subfigure}[b]{0.85\textwidth}
       \includegraphics[width=\textwidth]{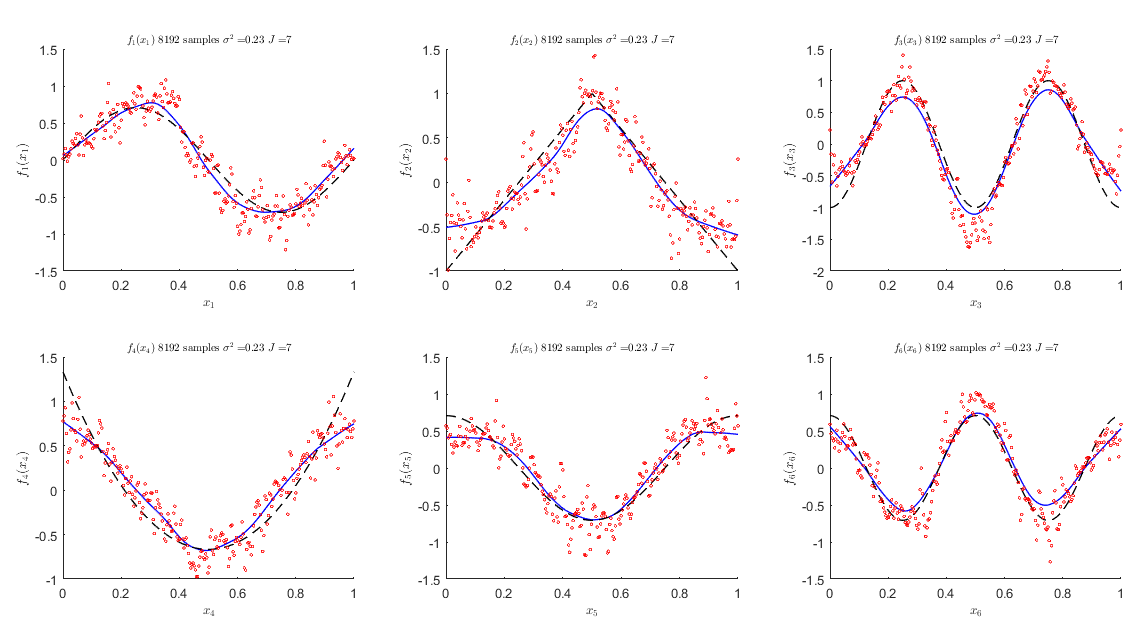}
       \caption{}
       \label{fig:FunctEst8192BetaDesign_Efro}
   \end{subfigure}\vfill
   \begin{subfigure}[b]{0.85\textwidth}
       \includegraphics[width=\textwidth]{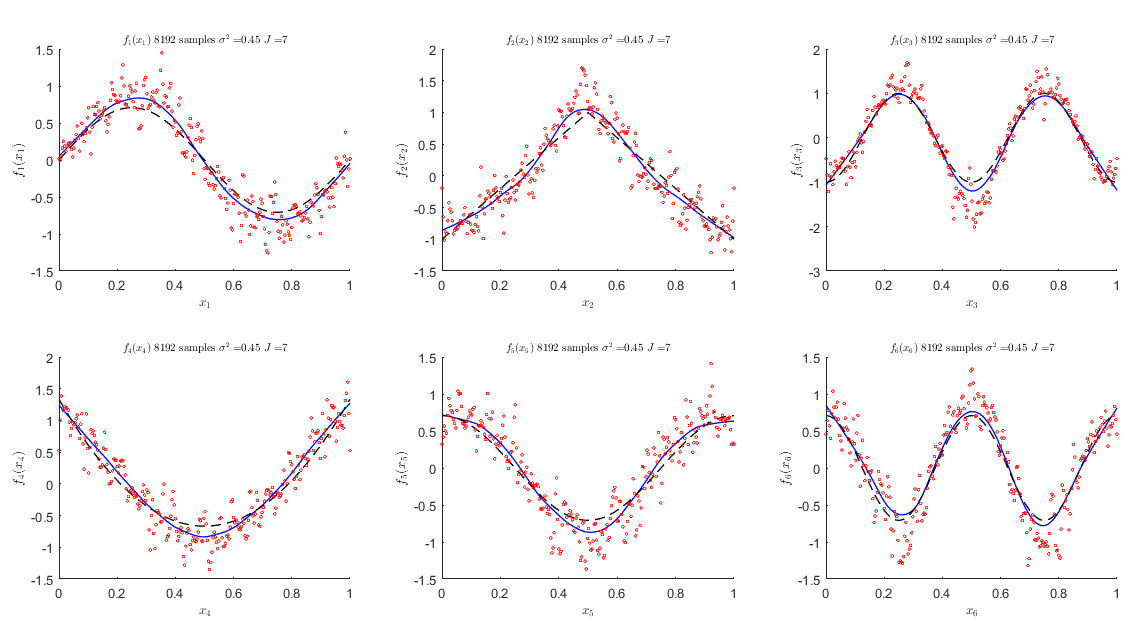}
       \caption{}
       \label{fig:FunctEst8192UnifDesign_Efro}
    \end{subfigure}
\caption{Functions estimation for (a) $Beta(\frac{3}{2},\frac{3}{2})$  and (b) $\mathcal{U}(0,1)$  designs, for $n=8192$ samples. In red, the estimated
function values at each sample point; In black-dashed lines, the actual function shape; In blue lines, the smoothed version of the function values using lowess smoother.}
\label{fig:EstFuncsEfro8192}
\end{figure}

\subsubsection*{Remarks and comments}
\begin{enumerate}[(i)]
\item Choice of bandwidth for the density estimator $\hat{h}_{n}(\textbf{x})$: During the implementation, we observed that results were highly sensitive to the choice of the bandwidth $\delta(n)$. We chose different values for a constant $K$ in a bandwidth of the form $\delta(n)=K\,n^{-\frac{1}{2+p}}$. Figures \ref{fig:FunctEst512BetaDesign_Efro}-\ref{fig:FunctEst8192UnifDesign_Efro} show results obtained using $K=3$.
\item Sample size effect: As can be observed in \ref{fig:FunctEst512BetaDesign_Efro}-\ref{fig:FunctEst8192UnifDesign_Efro}, both the bias and the variance of the estimated functions show a decreasing behavior as $n$ increases, which is consistent with theoretical results (\ref{eq:Risk4}), (\ref{eq:Risk5}) and (\ref{eq:Risk6}).
\item Shadowing effect of the constant $\beta_{0}$: In some experiments, when the constant $\beta_{0}$ was too large with respect to the function effects, we observed that the method recovered the marginal densities of each predictor instead of the unknown functions. This effect can be explained from the expressions for the calculation of the empirical wavelet coefficients $\hat{c}_{Jk}^{(l)}$. For this reason, we recommend standardizing the response from the observed sample before fitting the model.
\item Sensitivity of the model to different random designs: In the case of design distributions that have fast decaying tails, problems were observed when there was no sufficient information for the estimation of the empirical coefficients in regions with low concentration of samples. Indeed, extremely large empirical wavelet coefficients were obtained in those cases, inflating the bias in the estimation.
\item A possible remedial action for situation could be the use of the approach proposed in \cite{Pensky2001}, by thresholding the density estimates according to some probabilistic rule, avoiding those samples for which $\hat{h}_{n}(\textbf{x})$ is smaller than a suitably defined $\lambda_{n}>0$.
\end{enumerate}

\subsection{Conclusions and Discussion}

This paper introduced a wavelet-based method for the non-parametric estimation and prediction of non-linear additive regression models. Our estimator is based on data-driven wavelet coefficients computed using a locally weighted average of the observed samples, with weights defined by scaling functions obtained from an orthonormal periodic wavelet basis and a non-parametric density estimator $\hat{h}_{n}$. For this estimator, we showed mean-square consistency and illustrated practical results using theoretical simulations. In addition, we provided convergence rates and optimal choices for the tuning parameters for the algorithm implementation.

\medskip
As was seen in the sequel, the proposed estimator is completely data driven with only a few parameters of choice by the user (i.e. bandwidth $\delta(n)$, multiresolution index $J(n)$ and wavelet filter). Indeed, the nature of the estimator allows a block-matrix based implementation that introduces computational speed and makes the estimator suitable for real-life applications. In our implementation, Daubechies-Lagarias's algorithm was used to evaluate the scaling functions $\phi_{Jk}^{per}$ at the observed sample points $X_{ij}$.

\medskip
Furthermore, we tested our method using different exemplary baseline functions and two random designs via a theoretical simulation study. In our experiments, the proposed method showed good performance identifying the unknown functions in the model, even though it suffers from the "curse of dimensionality"; Also, we observed that the estimator behaves accordingly to the large properties behavior that were theoretically shown, which is an important feature for real-life applications.

\medskip
In terms of some of the drawbacks, we can mention that our method does not offer automatic variable selection; however, this could be implemented by the thresholding the obtained empirical wavelet coefficients in a post-estimation stage or by simple inspection, since a function that is zero over [0,1] maps to zero in the wavelet projection. Similarly, the proposed estimator was observed to be highly sensitive to the bandwidth choice $\delta(n)$, consequently, the use of cross-validation during the estimation stage might be helpful to improve the accuracy of results.

\medskip

Finally, in those design regions were the number of observed samples is small it is possible to obtain abnormaly large wavelet coefficients; also as a result of the use of periodic wavelets, some problems may arise at the boundaries of the support for each function. Nonetheless, this can be fixed: using the idea developed by Pensky and Vidakovic (2001) \cite{Pensky2001}, it is possible to avoid those samples that are associated with too-small density estimates $\hat{h}_{n}$, stabilizing the estimated wavelet coefficients and reducing the estimator bias.

\medskip

Based on out theoretical analysis and preliminary experiments, we can argue that our proposed method exhibits good statistical properties and is relatively easy to implement, which constitutes a good contribution in the statistical modeling field and in particular, in the analysis of the non-linear Additive regression models.

\newpage
\section*{References}
\bibliography{Refpaper3}
\bibliographystyle{unsrt}

\newpage

\appendix

\section{Proof of $\int_{0}^{1}\phi_{jk}^{per}(x)dx=2^{-\frac{j}{2}}$.} \label{proof:periodic}

For $j \leq 0$, the Strang-Fix condition (see \cite{Morettin2017}) gives  $\phi_{jk}(x) \equiv 2^{-j/2}$, so the claim is trivial. In the case of $j > 0$, it follows:
\begin{eqnarray}
\nonumber
\int_0^1 \phi_{jk}^{per} (x) dx & = & \sum_{m \in Z} \int_0^1 \phi_{jk}(x + m) dx \\
\nonumber
& = & \sum_{m \in Z} \int_0^1 2^{j/2} \phi(2^j (x + m) - k ) dx \\
\nonumber
&   &    ~~~[2^j (x + m) = t] \\
\nonumber
& = &  \sum_{m \in Z} \int_{m 2^j}^{(m+1) 2^j}  2^{j/2} 2^{-j}\phi(t - k ) dt \\
& = & 2^{-j/2} \int_R \phi(t - k) dt = 2^{-j/2}\,, \label{eq:proof1}
\end{eqnarray}
which shows the desired result.
\newpage

\section{Important results from Multivariate Taylor Series expansion.} \label{der:Taylor}

In this section we provide definitions and results that will be needed for the derivation of the density estimator $\hat{h}_{n}(\textbf{x})$ properties.

\medskip

Define $\boldsymbol{\alpha}:=(\alpha_{1},...,\alpha_{p})$, $\boldsymbol{\beta}:=(\beta_{1},...,\beta_{p})$, $|\boldsymbol{\alpha}|:=\sum_{j=1}^{p}\alpha_{j}$,
$|\boldsymbol{\beta}|:=\sum_{j=1}^{p}\beta_{j}$ and $\boldsymbol{\alpha}!=\prod_{j=1}^{p}\alpha_{j}!$. Similarly, let:
\begin{eqnarray}
\textbf{x}^{\boldsymbol{\alpha}} &:=& \prod_{j=1}^{p}x_{j}^{\alpha_{j}}, \,\, \textbf{x}\in \mathbb{R}^{p}\,, \label{eq:taylor1}\\
\partial^{\boldsymbol{\alpha}}f &:=& \partial_{1}^{\alpha_{1}}\cdot...\cdot\partial_{p}^{\alpha_{p}}f =
\frac{\partial^{|\boldsymbol{\alpha}|}f}{\partial x_{1}^{\alpha_{1}}\cdot...\cdot\partial x_{p}^{\alpha_{p}}}\,. \label{eq:taylor2}
\end{eqnarray}

\medskip

From the multinomial theorem,  it follows that for any $\textbf{x} \in \mathbb{R}^{p}$, and any integer $k>0$:

\begin{eqnarray}
\nonumber
|\textbf{x}|^{k}&=&\sum_{\alpha_{1}}\sum_{\alpha_{2}}\cdot...\cdot\sum_{\alpha_{p}}\frac{k!}{\alpha_{1}!\cdot...\cdot\alpha_{p}!}
x_{1}^{\alpha_{1}}\cdot...\cdot x_{p}^{\alpha_{p}}, \,\,s.t. \,\,|\boldsymbol{\alpha}|=k\,,\\
 &=& \sum_{|\boldsymbol{\alpha}|=k} \frac{k!}{\boldsymbol{\alpha}!}\textbf{x}^{\boldsymbol{\alpha}}\,. \label{eq:taylor3}
\end{eqnarray}
\medskip

Now, suppose a function $f:\mathbb{R}^{p}\rightarrow\mathbb{R}$, such that $f\in \mathbb{C}^{k}$ on a convex open set $\mathbb{S}\subset \mathbb{R}^{p}$. We are interested in the Taylor series expansion of $f(\textbf{x})$ around a point $\textbf{x}_{0}\in \mathbf{S}$.

\medskip

If we look at the behavior of $f()$ over the points that are in the line between $\textbf{x}$ and $\textbf{x}_{0}$, it follows that any of those points $\textbf{x}^{*}$ can be contained in a set defined as:

\begin{equation}
\nonumber
 L(\textbf{x},\textbf{x}_{0})=\left\{\textbf{x}^{*} \in \mathbf{S}
\,\,\,s.t.\, \forall t \in [0,1] \, \textbf{x}^{*}=\textbf{x}_{0}+t(\textbf{x}-\textbf{x}_{0}) \right\}\,.
\end{equation}

\medskip
Using the last definition, we have that $\forall \textbf{x}\in L(\textbf{x},\textbf{x}_{0})$, $f(\textbf{x}^{*})=f(\textbf{x}_{0}+t(\textbf{x}-\textbf{x}_{0}))=g(t)$. Define
$\textbf{v}=\textbf{x}-\textbf{x}_{0}$, therefore, for $1\leq l \leq k$, it follows:

\begin{equation}
\nonumber
g^{(l)}(t)=(\textbf{v}\bullet\nabla)^{l}\cdot f(\textbf{x}_{0}+t\cdot\textbf{v})\,,
\end{equation}
where
\begin{eqnarray}
\nonumber
(\textbf{v}\bullet\nabla)^{(l)}f&=&(v_{1}\frac{\partial}{\partial x_{1}}+...+v_{p}\frac{\partial}{\partial x_{p}})^{l}f\,, \\
\nonumber
&=& \sum_{|\boldsymbol{\alpha}|=l}\frac{l!}{\boldsymbol{\alpha}!}v_{1}^{\alpha_{1}}\cdot...\cdot v_{p}^{\alpha_{p}}\frac{\partial^{\alpha_{1}}}{\partial x_{1}^{\alpha_{1}}}\cdot...\cdot \frac{\partial^{\alpha_{p}}}{\partial x_{p}^{\alpha_{p}}}f \,,\\
&=& \sum_{|\boldsymbol{\alpha}|=l}\frac{l!}{\boldsymbol{\alpha}!}v_{1}^{\alpha_{1}}\cdot...\cdot v_{p}^{\alpha_{p}}\partial^{\boldsymbol{\alpha}}f\,.
\end{eqnarray}
If we now make a Taylor series expansion of $g(t)$ around a point $t_{0}$, for $\delta \in [t,t_{0}]$ it follows:

\begin{equation}
\nonumber
g(t)=\sum_{l=0}^{k-1}\frac{g^{(l)}(t_{0})}{l!}(t-t_{0})^{l}+\frac{g^{(k)}(\delta)(t-t_{0})^{k}}{k!}
\end{equation}
Letting $t_{0}\rightarrow0$ and $t\rightarrow1$, we have that $g^{(l)}(t_{0})\rightarrow \sum_{|\boldsymbol{\alpha}|=l}\frac{l!}{\boldsymbol{\alpha}!}v_{1}^{\alpha_{1}}\cdot...\cdot v_{p}^{\alpha_{p}}\partial^{\boldsymbol{\alpha}}f(\textbf{x}_{0})$ and $g(t)\rightarrow f(\textbf{x})$.

\medskip

Therefore, the Taylor series expansion of $f$ around $\textbf{x}_{0}$ is given by:

\begin{equation} \label{eq:taylor4}
f(\textbf{x})=\sum_{l=0}^{k-1}\frac{(\textbf{v}\bullet\nabla)^{(l)}f(\textbf{x}_{0})}{l!}+\frac{(\textbf{v}\bullet\nabla)^{(k)}f(\textbf{x}_{0}+\delta\textbf{v})}{k!}\,.
\end{equation}
Define the Taylor series expansion of $f()$ around $\textbf{x}_{0}$ of order $k$ and its remainder term as as:
\begin{eqnarray}
\nonumber
f_{\textbf{x}_{0},k}(\textbf{x})&=&\sum_{l=0}^{k-1}\frac{(\textbf{v}\bullet\nabla)^{(l)}f(\textbf{x}_{0})}{l!}\,, \\
\nonumber
R_{\textbf{x}_{0},k}(\textbf{v})&=&\frac{(\textbf{v}\bullet\nabla)^{(k)}f(\textbf{x}_{0}+\delta\textbf{v})}{k!}\,.
\end{eqnarray}.
Then, by Taylor's theorem and (\ref{eq:taylor3}), it follows:
\begin{equation}\label{eq:taylor5}
|R_{\textbf{x}_{0},k}(\textbf{v})|\leq \frac{M_{h}}{(k+1)!}||\textbf{v}||_{1}^{(k+1)}\,,
\end{equation}
provided assumption (\textbf{A4}) holds. Finally, from results (\ref{eq:taylor4}) and (\ref{eq:taylor5}), it follows that:
\begin{equation}
f(\textbf{x})-f_{\textbf{x}_{0},k}(\textbf{x})=R_{\textbf{x}_{0},k}(\textbf{v})\,. \label{eq:k.1}
\end{equation}

\section{Consistency of the Kernel density estimator.} \label{proof:kernelconsistency}

In this section, we provide an overview of the asymptotic properties of the density estimator $\hat{h}_{n}()$, which are needed later to show the consistency of the
estimates $\hat{\beta}_{0}$ and $\hat{c}_{Jk}^{(l)}$. See \cite{Wied2010} for a detailed discussion of the Kernel Density estimator properties.

\medskip

Consider a kernel-type density estimator given by:

\begin{equation}\label{eq:4.1}
\hat{h}_{n}(\textbf{x})=\frac{1}{n}\sum_{i=1}^{n}\frac{1}{\delta^{p}}K\left(\frac{\textbf{x}-\textbf{x}_{i}}{\delta} \right)\,,
\end{equation}
where $\frac{1}{\delta^{p}}K\left(\frac{\textbf{x}-\textbf{x}_{i}}{\delta} \right):=K_{\delta}(\textbf{x},\textbf{x}_{i})$ and $\delta=\delta(n)>0$ is a proper bandwidth, and $K(\textbf{x})> 0$ is the kernel function. This last condition guarantees that $\hat{h}_{n}(\textbf{x})$ is non-negative and
continuous as a finite sum of positive and continuous functions.

\medskip

From (\ref{eq:3.9}) and (\ref{eq:3.12}) it is clear that we need a kernel function such that $\hat{h}_{n}(\textbf{x})>0$ and bounded in the support of $h()$. Assume that the chosen kernel satisfies:
\medskip
\paragraph{(Ak1)} $K(\textbf{x})$ is real-valued, Borel measurable function with $||K||_{\infty}<\infty$.
\paragraph{(Ak2)} $K(\textbf{x})$ has $\beta-1$ ($\beta\geq2$) vanishing moments, i.e. $\int K(\textbf{v})||\textbf{v}||_{1}^{s}d\textbf{v}=0, \,\, s=1,...,\beta-1$.
\paragraph{(Ak3)} $K(\textbf{x})$ belongs to $\mathbb{L}_{2}(\mathbb{R}^{p})$.
\paragraph{(Ak4)} $K(\textbf{x})$ satisfies $\int K(\textbf{v})d\textbf{v}=1$ and $\int K(\textbf{v})||\textbf{v}||_{1}^{\beta}d\textbf{v}=M_{k,\beta}<\infty$.
\paragraph{(Ak5)} $\mathop{\sup}\limits_{\textbf{x},\textbf{y}\in[0,1]^{p}}\left|K_{\delta}(\textbf{x},\textbf{y})\right|  \leq C_{1}\delta^{-p}$, for $\delta=\delta(n)>0$, $C_{1}>0$.
\paragraph{(Ak6)} $\mathop{\sup}\limits_{\textbf{x}\in[0,1]^{p}}\mathbb{E}\left[\left(K_{\delta}^{2}(\textbf{x},\textbf{x}_{i})\right)\right]  \leq C_{2}\delta^{-p}$, for $\delta=\delta(n)>0$, $C_{1}>0$, $C_{2}>0$.
\medskip
\medskip

\subsubsection*{Proposition 1} \label{prop1}
Consider a kernel that satisfies (\textbf{Ak1})-(\textbf{Ak6}) and a random variable \textbf{X} defined on a probability space $(\Omega,\Im,\mathbb{P})$ with density $h()$. Assume (\textbf{A1}) and (\textbf{A5}) are satisfied, then $\hat{h}_{n}()$ is consistent, provided $n\delta^{p}\rightarrow\infty$ and $\delta^{p}\rightarrow0$ as $n\rightarrow\infty$.
\medskip

This means that $\forall \textbf{x}\in [0,1]^{p}$ for which $\mathbb{P}\left\{\omega\in \Omega \,|\, \textbf{X}(\omega)=\textbf{x} \right\}>0$, it follows:

\begin{equation}\label{eq:4.2}
\hat{h}_{n}(\textbf{x})\mathop{\rightarrow}\limits^{\mathbb{P}}h(\textbf{x})
\end{equation}

\subsubsection*{Proof}

Consider an iid sample $\left\{y_{i},\textbf{x}_{i} \right\}_{i=1}^{n}$. It follows that the expectation of the density estimator (\ref{eq:4.1}) takes the form:
\begin{equation}
\nonumber 
\mathbb{E}[\hat{h}_{n}(\textbf{x})] = \int K(\textbf{v})h(\textbf{x}+\delta\textbf{v})d\textbf{v}
\end{equation}
If we subtract $h(\textbf{x})$ from the above expression, we get:

\begin{eqnarray}
\nonumber 
\mathbb{E}[\hat{h}_{n}(\textbf{x})-h(\textbf{x})] &=& \int K(\textbf{v})\left[h(\textbf{x}+\delta\textbf{v})-h(\textbf{x})\right]d\textbf{v}\,, \\
\nonumber
 &=& \int K(\textbf{v})\left[h(\textbf{x}+\delta\textbf{v})-h_{\textbf{x},\beta}(\textbf{x}+\delta\textbf{v})+h_{\textbf{x},\beta}(\textbf{x}+\delta\textbf{v})-h(\textbf{x})\right]d\textbf{v}\,, \\
 &=& \int K(\textbf{v})\left[h(\textbf{x}+\delta\textbf{v})-h_{\textbf{x},\beta}(\textbf{x}+\delta\textbf{v})\right]d\textbf{v}+
 \int  K(\textbf{v})\left[h_{\textbf{x},\beta}(\textbf{x}+\delta\textbf{v})-h(\textbf{x})\right]d\textbf{v}\,, \label{eq:k.2}
\end{eqnarray}
provided assumption (\textbf{Ak4}) holds.
\medskip

From (\ref{eq:taylor4}) that in the second term of (\ref{eq:k.2}): $h(\textbf{x}+\delta\textbf{v})_{\textbf{x},\beta}-h(\textbf{x})=\sum_{l=1}^{k-1}\frac{(\textbf{v}\bullet\nabla)^{(l)}f(\textbf{x}_{0})}{l!}$. Morover, by assumption (\textbf{Ak2}):

\begin{equation}
\int  K(\textbf{v})\left[h_{\textbf{x},\beta}(\textbf{x}+\delta\textbf{v})-h(\textbf{x})\right]d\textbf{v}=0\,.\label{eq:k.3}
\end{equation}
\medskip

Similarly, the first term of the rhs of (\ref{eq:k.2}) can be expressed as: $h(\textbf{x}+\delta\textbf{v})-h_{\textbf{x},\beta}(\textbf{x}+\delta\textbf{v})=R_{\textbf{x},\beta}(\delta\textbf{v})$, provided (\ref{eq:k.1}).  Therefore, from (\ref{eq:taylor5}), it follows:

\begin{eqnarray}
\nonumber
\mathbb{E}[\hat{h}_{n}(\textbf{x})-h(\textbf{x})] &=& \int K(\textbf{v})R_{\textbf{x},\beta}(\delta\textbf{v})d\textbf{v}\,, \\
\nonumber
|\mathbb{E}[\hat{h}_{n}(\textbf{x})-h(\textbf{x})]| &\leq & \int K(\textbf{v})|R_{\textbf{x},\beta}(\delta\textbf{v})|d\textbf{v}\,, \\
\nonumber
 &\leq & \frac{M_{h}\delta^{\beta}}{\beta!}\int K(\textbf{v})||\textbf{v}||_{1}^{\beta}d\textbf{v}\,, \\
|bias(\hat{h}_{n})| &\leq & C(h,\beta)\delta^{\beta}\,, \label{eq:k.4}
\end{eqnarray}
where $C(h,\beta)=\frac{M_{h}M_{k,\beta}}{\beta!}$. Also, from the last set of equations, it is possible to obtain:

\begin{equation}\label{eq:k.4b}
\mathop{\sup}\limits_{\textbf{x}\in[0,1]^{p}}\left|\mathbb{E}[\hat{h}_{n}(\textbf{x})-h(\textbf{x})]\right|  \leq C(h,\beta)\delta^{\beta}\,.
\end{equation}
\medskip

Now, for a fixed $\textbf{x}$, the variance of $\hat{h}_{n}(\textbf{x})$, can be expressed and bounded as follows:

\begin{eqnarray}
\nonumber 
  Var\left( \hat{h}_{n}(\textbf{x})\right) &=& \frac{1}{n\delta^{2p}}Var\left(K\left(\frac{\textbf{x}-\textbf{X}_{1}}{\delta}\right) \right)\,, \\
\nonumber
   & \leq & \frac{1}{n\delta^{2p}}\mathbb{E}\left[K\left(\frac{\textbf{x}-\textbf{X}_{1}}{\delta}\right)^{2} \right]\,, \\
\nonumber
   & \leq & \frac{1}{n\delta^{p}}\int K(\textbf{v})^{2}h(\textbf{x}+\delta\textbf{v})d\textbf{v}\,, \\
   & \leq & \frac{M\cdot C}{n\delta^{p}}\,, \\
\mathop{\sup}\limits_{\textbf{x}\in[0,1]^{p}}\mathbb{E}\left[\left(\hat{h}_{n}(\textbf{x})-h(\textbf{x})\right)^{2}\right] & \leq & \frac{M\cdot C}{n\delta^{p}}\,,  \label{eq:k.5}
\end{eqnarray}
provided assumptions (\textbf{A6}) and (\textbf{Ak3}) hold, for $C=\int K(\textbf{v})^{2}d\textbf{v}$.

\medskip

From the above results, it is possible to express the $\mathbb{L}_{2}$ risk of the estimator $\hat{h}_{n}(\textbf{x})$ as:

\begin{equation}
\nonumber
\mathbf{R}\left(\hat{h}_{n},h\right)=Var\left( \hat{h}_{n}(\textbf{x})\right)+bias(\hat{h}_{n}(\textbf{x}))^{2}\,.
\end{equation}
Using results (\ref{eq:k.4}) and (\ref{eq:k.5}), we get that:

\begin{equation}\label{eq:k.6}
\mathbf{R}\left(\hat{h}_{n},h\right) \leq \frac{M\cdot C}{n\delta^{p}}+C(h,\beta)^{2}\delta^{2\beta}
\end{equation}
Clearly, as $n\rightarrow\infty$, if $n\delta^{p}\rightarrow\infty$ and $\delta^{p}\rightarrow0$, it follows that $\mathbf{R}\left(\hat{h}_{n},h\right)\rightarrow0$. Therefore, $\hat{h}_{n}(\textbf{x})$ is mean-square consistent, which automatically implies:

\begin{equation}
\nonumber
\hat{h}_{n}(\textbf{x})\mathop{\rightarrow}\limits^{\mathbb{P}}h(\textbf{x})\,.
\end{equation}
\medskip

If we ignore the constants (with respect to $n$) in (\ref{eq:k.6}), it is possible to show that the bandwidth $\delta(n)$ that minimizes  $\mathbf{R}\left(\hat{h}_{n},h\right)$ is given by $\delta^{*}\sim n^{-\frac{1}{2\beta+p}}$ (up to a constant) and thus, $ \mathbf{R}\left(\hat{h}_{n},h\right)^{*}\geq C\cdot n^{-\frac{2\beta}{2\beta+p}}$. Similarly, under this optimal bandwidth, we have that (\ref{eq:k.5}) becomes:

\begin{equation}\label{eq:k.7}
\mathop{\sup}\limits_{\textbf{x}\in[0,1]^{p}}\mathbb{E}\left[\left(\hat{h}_{n}(\textbf{x})-h(\textbf{x})\right)^{2}\right] \leq M\cdot C n^{-\frac{2\beta}{2\beta+p}}\,.
\end{equation}
\newpage

\section{Derivation of an upper bound for $\mathbb{E}\left[\left(\frac{Y\phi_{Jk}^{per}(X_{l})}{h(\textbf{X})} \right)^{2} \right]$.} \label{proof:ExpUpper}

Consider a sequence of constant positive piecewise functions $\left\{g_{b},\,b\geq1\right\}$ that satisfy:

\begin{enumerate}[(i)]
  \item $0<g_{b}(\textbf{x})\leq h(\textbf{x})$, $\forall\,\textbf{x}\in [0,1]^{p}$.
  \item $g_{b}(\textbf{x})\leq g_{b+1}(\textbf{x})$, $\forall\,\textbf{x}\in [0,1]^{p}$.
  \item $g_{b}(\textbf{x}) \uparrow h(\textbf{x})$ as $b\rightarrow \infty$.
\end{enumerate}
Define $g_{b}(\textbf{x})$ for $b\geq \lfloor \log_{2}\left(\frac{1}{\epsilon_{h}} \right)\rfloor$ as follows:

\begin{equation}
\nonumber
g_{b}(\textbf{x}) = \begin{cases}
            \frac{r}{2^{b}} & \quad \frac{r}{2^{b}} \leq h(\textbf{x})\leq \frac{r+1}{2^{b}} \quad r=1,...,b\cdot 2^{b}-1 \\
            b & \quad h(\textbf{x}) > b
        \end{cases}
\end{equation}
Therefore, we can express $g_{b}(\textbf{x})$ as:

\begin{equation}\label{eq:ExpUpper.1}
g_{b}(\textbf{x}) = \sum_{r=1}^{b\cdot 2^{b}-1}\left(\frac{r}{2^{b}}\right)\mathbf{1}_{\left\{\textbf{x}:\,\frac{r}{2^{b}} \leq h(\textbf{x})\leq \frac{r+1}{2^{b}} \right\}}+
b\cdot\mathbf{1}_{\left\{\textbf{x}:\,h(\textbf{x}) > b\right\} }\,.
\end{equation}
From (\ref{eq:ExpUpper.1}), for a fixed $b$ define:
\begin{eqnarray}
\nonumber
\Omega_{rb}&=&\left\{\textbf{x}:\,\frac{r}{2^{b}} \leq h(\textbf{x})\leq \frac{r+1}{2^{b}}\right\}\,, r=1,...,b\cdot 2^{b}-1\,,\\
\nonumber
\Omega_{b}&=&\left\{\textbf{x}:\, h(\textbf{x})>b\right\}\,.
\end{eqnarray}
This partitions the support of the random vector $\textbf{X}$ into $b\cdot 2^{b}$ disjoints subsets for which $\bigcup_{r=1}^{b\cdot2^{b}-1}\left\{\Omega_{rb}\right\}\bigcup\left\{\Omega_{b}\right\}=[0,1]^{p}$. Similarly, the sequence of functions $\left\{g_{b},\,b\geq1\right\}$ approximate $h(\textbf{x})$ from below, in a quantization fashion. Therefore:

\begin{equation}
\nonumber
\mathbb{E}\left[\left(\frac{Y\phi_{Jk}^{per}(X_{l})}{h(\textbf{X})} \right)^{2} \right]=\sum_{r=1}^{b\cdot 2^{b}-1}\mathbb{E}\left[\left(\frac{Y\phi_{Jk}^{per}(X_{l})}{h(\textbf{X})} \right)^{2}\mathbf{1}_{\left\{\textbf{x}:\,\frac{r}{2^{b}} \leq h(\textbf{x})\leq \frac{r+1}{2^{b}} \right\}} \right]+ \mathbb{E}\left[\left(\frac{Y\phi_{Jk}^{per}(X_{l})}{h(\textbf{X})} \right)^{2}\mathbf{1}_{\left\{\textbf{x}: h(\textbf{x})> b \right\}}\right]\,,
\end{equation}
\begin{equation}
\nonumber
\resizebox{.88 \linewidth}{!}
{
$\mathbb{E}\left[\left(\frac{Y\phi_{Jk}^{per}(X_{l})}{h(\textbf{X})} \right)^{2} \right] \leq  \left(\left(p\cdot M_{f}+|\beta_{0}|\right)^{2}+\sigma^{2}\right)\left(\sum_{r=1}^{b\cdot 2^{b}-1}\mathbb{E}\left[\frac{\phi_{Jk}^{per}(X_{l})^{2}\mathbf{1}_{\left\{\textbf{x}:\,\frac{r}{2^{b}} \leq h(\textbf{x})\leq \frac{r+1}{2^{b}} \right\}}}{h(\textbf{X})^{2}}\right]+\mathbb{E}\left[\frac{\phi_{Jk}^{per}(X_{l})^{2}\mathbf{1}_{\left\{\textbf{x}:\, h(\textbf{x})>b \right\}}}{h(\textbf{X})^{2}}\right]\right)$}\,,
\end{equation}
\begin{eqnarray}
\nonumber
\mathbb{E}\left[\left(\frac{Y\phi_{Jk}^{per}(X_{l})}{h(\textbf{X})} \right)^{2} \right] &\leq & \left(\left(p\cdot M_{f}+|\beta_{0}|\right)^{2}+\sigma^{2}\right)\left(\sum_{r=1}^{b\cdot 2^{b}-1}\int_{\Omega_{rb}}\frac{\phi_{Jk}^{per}(X_{l})^{2} }{h(\textbf{X})}d\textbf{x} + \int_{\Omega_{b}}\frac{\phi_{Jk}^{per}(X_{l})^{2} }{h(\textbf{X})}d\textbf{x}\right)\,, \\
\nonumber
&\leq & \left(\left(p\cdot M_{f}+|\beta_{0}|\right)^{2}+\sigma^{2}\right)\left(\sum_{r=1}^{b\cdot 2^{b}-1}\int_{\Omega_{rb}}\frac{\phi_{Jk}^{per}(X_{l})^{2} }{g_{b}(\textbf{X})}d\textbf{x} + \int_{\Omega_{b}}\frac{\phi_{Jk}^{per}(X_{l})^{2} }{g_{b}(\textbf{X})}d\textbf{x}\right)\,, \\
\nonumber
&\leq & \left(\left(p\cdot M_{f}+|\beta_{0}|\right)^{2}+\sigma^{2}\right)\left(\sum_{r=1}^{b\cdot 2^{b}-1}\frac{2^{b}}{r}\int_{\Omega_{rb}}\phi_{Jk}^{per}(X_{l})^{2}d\textbf{x} +\frac{1}{b}\int_{\Omega_{b}}\phi_{Jk}^{per}(X_{l})^{2}d\textbf{x}\right)\,,\\
\nonumber
&\leq & \left(\left(p\cdot M_{f}+|\beta_{0}|\right)^{2}+\sigma^{2}\right)\left(2^{b}(b2^{b}-1)+\frac{1}{b} \right)\,, \\
\nonumber
&\leq & \left(\left(p\cdot M_{f}+|\beta_{0}|\right)^{2}+\sigma^{2}\right)\left\{\mathop{\inf}\limits_{b\geq \lfloor \log_{2}\left(\frac{1}{\epsilon_{h}} \right)\rfloor}\left(2^{b}(b2^{b}-1)+\frac{1}{b}\right) \right\} \,, \\
&\leq & \left(\left(p\cdot M_{f}+|\beta_{0}|\right)^{2}+\sigma^{2}\right)\left\{
 \frac{1}{\epsilon_{h}}\left(\lceil\log_{2}(\frac{1}{\epsilon_{h}})\rceil-1 \right)+\frac{1}{\lceil \log_{2}(\frac{1}{\epsilon_{h}})\rceil}\right\} \,, \label{eq:ExpUpper.2}
\end{eqnarray}
where the last result holds since the function $f(b)=2^{b}(b2^{b}-1)+\frac{1}{b}$ is strictly increasing in $b$ and $b\geq \lfloor \log_{2}\left(\frac{1}{\epsilon_{h}} \right)\rfloor$.

\subsection*{Remarks}
Note that this bound could be further improved if instead of piecewise constant functions, we use a different approximation technique. Nonetheless, obtaining tight bounds is not the intention of this derivations, but instead showing that the second moment of the random variable $\frac{Y\phi_{Jk}^{per}(X_{l})}{h(\textbf{X})}$ is bounded under suitable conditions.

\newpage

\section{Asymptotic correlation between $\frac{Y_{i}\phi_{Jk}^{per}(X_{il})}{\hat{h}_{n}(\textbf{X}_{i})}$ and $\hat{\beta}_{0}$.} \label{proof:asympUncorrelation}

Similarly as for $V_{c1}$ in (\ref{eq:4.17}), consider the asymptotic behavior of $V_{c3}=Cov\left(\frac{1}{n}\sum_{i=1}^{n}\frac{Y_{i}\phi_{Jk}^{per}(X_{il})}{\hat{h}_{n}(\textbf{X}_{i})}\,,\,2^{-\frac{J}{2}}\hat{\beta}_{0}\right)$ assuming conditions (\textbf{A1})-(\textbf{A5}) and (\textbf{Ak1})-(\textbf{Ak6}) hold. Using the covariance properties and the iid sample $\{y_{i}=f(\textbf{x}_{i}),\textbf{x}_{i} \}_{i=1}^{n}$, it follows:

\begin{equation}\label{eq:asympUncorrelation.1}
  V_{c3} = \frac{2^{-\frac{J}{2}}}{n^{2}}\left\{\sum_{i=1}^{n}Cov\left(\frac{Y_{i}\phi_{Jk}^{per}(X_{il})}{\hat{h}_{n}(\textbf{X}_{i})}\,,\,\frac{Y_{i}}{\hat{h}_{n}(\textbf{X}_{i})} \right)+
  \mathop{\sum_{i=1}^{n}\sum_{j=1}^{n}}\limits_{i\neq j}Cov\left(\frac{Y_{i}\phi_{Jk}^{per}(X_{il})}{\hat{h}_{n}(\textbf{X}_{i})}\,,\,\frac{Y_{j}}{\hat{h}_{n}(\textbf{X}_{j})} \right)\right\}\,.
\end{equation}
\subsubsection*{Case $i=j$}

We have for $i=j$, $i=1,...,n$:

\begin{eqnarray}
\nonumber 
 Cov\left(\frac{Y_{i}\phi_{Jk}^{per}(X_{il})}{\hat{h}_{n}(\textbf{X}_{i})}\,,\,\frac{Y_{i}}{\hat{h}_{n}(\textbf{X}_{i})} \right)  &=& \mathbb{E}\left[\frac{Y^{2}\phi_{Jk}^{per}(X_{l})}{\hat{h}_{n}(\textbf{X})^{2}}\right] - \mathbb{E}\left[\frac{Y\phi_{Jk}^{per}(X_{l})}{\hat{h}_{n}(\textbf{X})}\right]\mathbb{E}\left[\frac{Y}{\hat{h}_{n}(\textbf{X})} \right]\,.
\end{eqnarray}
Using conditional expectation in the same way as in \ref{eq:4.4} and applying dominated convergence, it follows:

\begin{equation}\label{eq:asympUncorrelation.2}
Cov\left(\frac{Y_{i}\phi_{Jk}^{per}(X_{il})}{\hat{h}_{n}(\textbf{X}_{i})}\,,\,\frac{Y_{i}}{\hat{h}_{n}(\textbf{X}_{i})} \right) \mathop{\rightarrow}\limits_{n\rightarrow\infty}
\mathbb{E}\left[\frac{Y^{2}\phi_{Jk}^{per}(X_{l})}{h(\textbf{X})^{2}}\right]-\beta_{0}\mathbb{E}\left[\frac{Y\phi_{Jk}^{per}(X_{l})}{h(\textbf{X})}\right]\,.
\end{equation}
\subsubsection*{Case $i\neq j$}

For $i\neq j$, $i,j=1,...,n$, it is possible to obtain:
\begin{eqnarray}
\nonumber 
 Cov\left(\frac{Y_{i}\phi_{Jk}^{per}(X_{il})}{\hat{h}_{n}(\textbf{X}_{i})}\,,\,\frac{Y_{j}}{\hat{h}_{n}(\textbf{X}_{j})} \right)  &=& \mathbb{E}\left[\frac{Y_{i}Y_{j}\phi_{Jk}^{per}(X_{il})}{\hat{h}_{n}(\textbf{X}_{i})\hat{h}_{n}(\textbf{X}_{j})}\right] - \mathbb{E}\left[\frac{Y\phi_{Jk}^{per}(X_{l})}{\hat{h}_{n}(\textbf{X})}\right]\mathbb{E}\left[\frac{Y}{\hat{h}_{n}(\textbf{X})} \right]\,.
\end{eqnarray}
From the definition of $\hat{h}_{n}(\textbf{X})$ in (\ref{eq:4.1}), it follows:
\begin{eqnarray}
\nonumber
\hat{h}_{n}(\textbf{X}_{i}) &=& \frac{K(\textbf{0})}{n\delta^{p}}+\frac{n-1}{n}\hat{h}_{n-1}(\textbf{X}_{i})\,,
\end{eqnarray}
therefore, for $n$ sufficiently large:

\begin{equation}
\nonumber
\hat{h}_{n}(\textbf{X}_{i}) \approx \hat{h}_{n-1}^{(-i)}(\textbf{X}_{i})\,,
\end{equation}
provided $n\delta^{p}$ uniformly goes to $\infty$, where $\hat{h}_{n-1}^{(-i)}(\textbf{X}_{i})$ corresponds to the kernel density estimator computed without the $i-$th sample, evaluated at $\textbf{X}_{i}$.

\medskip

Let $\textbf{X}^{(-i,-j)}$ denote the sample $\left\{\textbf{X}_{1},...,\textbf{X}_{n}\right\}$ without $\textbf{X}_{i},\textbf{X}_{j}$. Therefore, using conditional expectation and for $n$ sufficiently large:

\begin{eqnarray}
\nonumber
\mathbb{E}\left[\frac{Y_{i}Y_{j}\phi_{Jk}^{per}(X_{il})}{\hat{h}_{n}(\textbf{X}_{i})\hat{h}_{n}(\textbf{X}_{j})}\right] &=& \mathbb{E}_{\textbf{X}_{i},\textbf{X}_{j}}\left[\mathbb{E}_{\textbf{X}^{(-i,-j)}|\textbf{X}_{i},\textbf{X}_{j}}\left[\frac{Y_{i}Y_{j}\phi_{Jk}^{per}(X_{il})}{\hat{h}_{n}(\textbf{X}_{i})\hat{h}_{n}(\textbf{X}_{j})} |\textbf{X}_{i},\textbf{X}_{j}\right] \right]\,, \\
\nonumber
&=& \mathbb{E}_{\textbf{X}_{i},\textbf{X}_{j}}\left[Y_{i}Y_{j}\phi_{Jk}^{per}(X_{il})\cdot\mathbb{E}_{\textbf{X}^{(-i,-j)}|\textbf{X}_{i},\textbf{X}_{j}}\left[\frac{1}{\hat{h}_{n}(\textbf{X}_{i})\hat{h}_{n}(\textbf{X}_{j})} |\textbf{X}_{i},\textbf{X}_{j}\right] \right]\,, \\
\nonumber
& \approx & \mathbb{E}_{\textbf{X}_{i},\textbf{X}_{j}}\left[Y_{i}Y_{j}\phi_{Jk}^{per}(X_{il})\cdot\mathbb{E}_{\textbf{X}^{(-i,-j)}|\textbf{X}_{i},\textbf{X}_{j}}\left[\frac{1}{\hat{h}_{n-1}^{(-i)}(\textbf{X}_{i})\hat{h}_{n-1}^{(-j)}(\textbf{X}_{j})} |\textbf{X}_{i},\textbf{X}_{j}\right] \right]\,.
\end{eqnarray}
Using the last result and dominated convergence, it follows:

\begin{eqnarray}
\nonumber
\mathbb{E}\left[\frac{Y_{i}Y_{j}\phi_{Jk}^{per}(X_{il})}{\hat{h}_{n}(\textbf{X}_{i})\hat{h}_{n}(\textbf{X}_{j})}\right] & \mathop{\rightarrow}\limits_{n\rightarrow\infty} & \mathbb{E}_{\textbf{X}_{i},\textbf{X}_{j}}\left[\frac{Y_{i}Y_{j}\phi_{Jk}^{per}(X_{il})}{h(\textbf{X}_{i})h(\textbf{X}_{j})}\right]\,, \\
&\mathop{\rightarrow}\limits_{n\rightarrow\infty} & \beta_{0}\cdot\mathbb{E}\left[\frac{Y\phi_{Jk}^{per}(X_{l})}{h(\textbf{X})} \right]\,, \label{eq:asympUncorrelation.3}
\end{eqnarray}
provided the iid condition of the observed sample. Finally,

\begin{eqnarray}
\nonumber
Cov\left(\frac{Y_{i}\phi_{Jk}^{per}(X_{il})}{\hat{h}_{n}(\textbf{X}_{i})}\,,\,\frac{Y_{j}}{\hat{h}_{n}(\textbf{X}_{j})} \right) &\mathop{\rightarrow}\limits_{n\rightarrow\infty} &
\beta_{0}\cdot\mathbb{E}\left[\frac{Y\phi_{Jk}^{per}(X_{l})}{h(\textbf{X})} \right]-\beta_{0}\cdot\mathbb{E}\left[\frac{Y\phi_{Jk}^{per}(X_{l})}{h(\textbf{X})} \right]\,, \\
&\mathop{\rightarrow}\limits_{n\rightarrow\infty} & 0 \,.\label{eq:asympUncorrelation.4}
\end{eqnarray}
\medskip

Therefore, using (\ref{eq:asympUncorrelation.2}) and (\ref{eq:asympUncorrelation.4}) in (\ref{eq:asympUncorrelation.1}), it follows:

\begin{eqnarray}
\nonumber 
  V_{c3} &=&  \frac{2^{-\frac{J}{2}}}{n^{2}}\left\{nCov\left(\frac{Y_{i}\phi_{Jk}^{per}(X_{il})}{\hat{h}_{n}(\textbf{X}_{i})}\,,\,\frac{Y_{i}}{\hat{h}_{n}(\textbf{X}_{i})} \right)+n(n-1)Cov\left(\frac{Y_{i}\phi_{Jk}^{per}(X_{il})}{\hat{h}_{n}(\textbf{X}_{i})}\,,\,\frac{Y_{j}}{\hat{h}_{n}(\textbf{X}_{j})} \right) \right\}\,, \\
\nonumber
   &=& 2^{-\frac{J}{2}}\left\{\frac{1}{n}Cov\left(\frac{Y_{i}\phi_{Jk}^{per}(X_{il})}{\hat{h}_{n}(\textbf{X}_{i})}\,,\,\frac{Y_{i}}{\hat{h}_{n}(\textbf{X}_{i})} \right)+\frac{n(n-1)}{n^{2}}Cov\left(\frac{Y_{i}\phi_{Jk}^{per}(X_{il})}{\hat{h}_{n}(\textbf{X}_{i})}\,,\,\frac{Y_{j}}{\hat{h}_{n}(\textbf{X}_{j})} \right) \right\}\,.
\end{eqnarray}
This last result implies:

\begin{equation}\label{eq:asympUncorrelation.5}
Cov\left(\frac{1}{n}\sum_{i=1}^{n}\frac{Y_{i}\phi_{Jk}^{per}(X_{il})}{\hat{h}_{n}(\textbf{X}_{i})}\,,\,2^{-\frac{J}{2}}\hat{\beta}_{0}\right)   \mathop{\rightarrow}\limits_{n\rightarrow\infty} 0\,.
\end{equation}
As a corollary, we can see that from (\ref{eq:asympUncorrelation.5}), it follows that $Cov\left(\hat{\beta}_{0}\,,\,\hat{c}_{Jk}^{(l)}\right)\mathop{\rightarrow}\limits_{n\rightarrow\infty} 0$. In fact, note that $Cov\left(\hat{\beta}_{0}\,,\,\hat{c}_{Jk}^{(l)}\right)$ can be expressed as:

\begin{equation}
\nonumber
Cov\left(\hat{\beta}_{0}\,,\,\hat{c}_{Jk}^{(l)}\right) = Cov\left(\hat{\beta}_{0}\,,\,\frac{1}{n}\sum_{i=1}^{n}\frac{Y_{i}\phi_{Jk}^{per}(X_{il})}{\hat{h}_{n}(\textbf{X}_{i})}\right) -2^{-\frac{J}{2}}Var\left(\hat{\beta}_{0}\right)\,.
\end{equation}
Therefore, from (\ref{eq:4.10}) and (\ref{eq:asympUncorrelation.5}), it is clear that $Cov\left(\hat{\beta}_{0}\,,\,\hat{c}_{Jk}^{(l)}\right)\mathop{\rightarrow}\limits_{n\rightarrow\infty} 0$ as desired.

\medskip

Finally, this asertion also implies that:

\begin{equation}\label{eq:asympUncorrelation.6}
Cov\left(\hat{\beta}_{0}\,,\,\sum_{l=1}^{p}\sum_{k=0}^{2^{J}-1}\hat{c}_{Jk}^{(l)}\phi_{Jk}^{per}(x_{l})\right) \mathop{\rightarrow}\limits_{n\rightarrow\infty} 0\,,
\end{equation}
by the properties of the covariance function.
\newpage

\section{Asymptotic convergence of $Cov\left(\hat{c}_{Jk}^{(l)}\,,\,\hat{c}_{Js}^{(l)} \right)$.} \label{proof:asympUncorrelationCoeff}

For any $s\neq k$, $s,k=0,...,2^{J}-1$ and fixed $J$, assuming conditions (\textbf{A1})-(\textbf{A5}) and (\textbf{Ak1})-(\textbf{Ak6}) hold, it follows:

\begin{eqnarray}
\nonumber
Cov\left(\hat{c}_{Jk}^{(l)}\,,\,\hat{c}_{Js}^{(l)} \right) &=& Cov\left(\frac{1}{n}\sum_{i=1}^{n}\frac{Y_{i}\phi_{Jk}^{per}(X_{il})}{\hat{h}_{n}(\textbf{X}_{i})}-2^{-\frac{J}{2}}\hat{\beta}_{0}\,,\,
\frac{1}{n}\sum_{i=1}^{n}\frac{Y_{i}\phi_{Js}^{per}(X_{il})}{\hat{h}_{n}(\textbf{X}_{i})}-2^{-\frac{J}{2}}\hat{\beta}_{0} \right)\,, \\
\nonumber
&=& \mathbb{E}\left[\frac{1}{n^{2}}\sum_{i=1}^{n}\sum_{j=1}^{n}\frac{Y_{i}Y_{j}\phi_{Jk}^{per}(X_{il})\phi_{Js}^{per}(X_{jl})}{\hat{h}_{n}(\textbf{X}_{i})\hat{h}_{n}(\textbf{X}_{j})} \right]-2^{-\frac{J}{2}}\mathbb{E}\left[\hat{\beta}_{0}\frac{1}{n}\sum_{i=1}^{n}\frac{Y_{i}\phi_{Jk}^{per}(X_{il})}{\hat{h}_{n}(\textbf{X}_{i})} \right] \\
\nonumber
& & -2^{-\frac{J}{2}}\mathbb{E}\left[\hat{\beta}_{0}\frac{1}{n}\sum_{i=1}^{n}\frac{Y_{i}\phi_{Js}^{per}(X_{il})}{\hat{h}_{n}(\textbf{X}_{i})} \right]+2^{-J}\mathbb{E}\left[\hat{\beta}_{0}^{2} \right]-\mathbb{E}\left[\hat{c}_{Jk}^{(l)}\right]\mathbb{E}\left[\hat{c}_{Js}^{(l)}\right]\,, \\
\nonumber
&=& \mathbb{E}\left[\frac{1}{n^{2}}\sum_{i=1}^{n}\sum_{j=1}^{n}\frac{Y_{i}Y_{j}\phi_{Jk}^{per}(X_{il})\phi_{Js}^{per}(X_{jl})}{\hat{h}_{n}(\textbf{X}_{i})\hat{h}_{n}(\textbf{X}_{j})}\right]
-2^{-\frac{J}{2}}Cov\left(\hat{\beta}_{0}\,,\,\frac{1}{n}\sum_{i=1}^{n}\frac{Y_{i}\phi_{Js}^{per}(X_{il})}{\hat{h}_{n}(\textbf{X}_{i})}\right)\\
\nonumber
& & -2^{-\frac{J}{2}}\mathbb{E}\left[\hat{\beta}_{0}\right]\mathbb{E}\left[\frac{1}{n}\sum_{i=1}^{n}\frac{Y_{i}\phi_{Js}^{per}(X_{il})}{\hat{h}_{n}(\textbf{X}_{i})} \right]-2^{-\frac{J}{2}}Cov\left(\hat{\beta}_{0}\,,\,\frac{1}{n}\sum_{i=1}^{n}\frac{Y_{i}\phi_{Jk}^{per}(X_{il})}{\hat{h}_{n}(\textbf{X}_{i})}\right) \\
\nonumber
& & - 2^{-\frac{J}{2}}\mathbb{E}\left[\hat{\beta}_{0}\right]\mathbb{E}\left[\frac{1}{n}\sum_{i=1}^{n}\frac{Y_{i}\phi_{Jk}^{per}(X_{il})}{\hat{h}_{n}(\textbf{X}_{i})} \right]+2^{-J}\mathbb{E}\left[\hat{\beta}_{0}^{2} \right]-\mathbb{E}\left[\hat{c}_{Jk}^{(l)}\right]\mathbb{E}\left[\hat{c}_{Js}^{(l)}\right]\,, \\
\nonumber
&=& \mathbb{E}\left[\frac{1}{n^{2}}\sum_{i=1}^{n}\sum_{j=1}^{n}\frac{Y_{i}Y_{j}\phi_{Jk}^{per}(X_{il})\phi_{Js}^{per}(X_{jl})}{\hat{h}_{n}(\textbf{X}_{i})\hat{h}_{n}(\textbf{X}_{j})}\right]
-2^{-\frac{J}{2}}Cov\left(\hat{\beta}_{0}\,,\,\frac{1}{n}\sum_{i=1}^{n}\frac{Y_{i}\phi_{Js}^{per}(X_{il})}{\hat{h}_{n}(\textbf{X}_{i})}\right) \\
\nonumber
& & 2^{-\frac{J}{2}}Cov\left(\hat{\beta}_{0}\,,\,\frac{1}{n}\sum_{i=1}^{n}\frac{Y_{i}\phi_{Jk}^{per}(X_{il})}{\hat{h}_{n}(\textbf{X}_{i})}\right)+2^{-J}Var\left(\hat{\beta}_{0}\right) \\
\nonumber
& & - \mathbb{E}\left[\frac{1}{n}\sum_{i=1}^{n}\frac{Y_{i}\phi_{Jk}^{per}(X_{il})}{\hat{h}_{n}(\textbf{X}_{i})}\right]
\mathbb{E}\left[\frac{1}{n}\sum_{i=1}^{n}\frac{Y_{i}\phi_{Js}^{per}(X_{il})}{\hat{h}_{n}(\textbf{X}_{i})}\right]\,.
\end{eqnarray}
Using the same argument that led to (\ref{eq:asympUncorrelation.3}), for $i\neq j$, it follows:

\begin{equation}
\nonumber
\mathbb{E}\left[\frac{Y_{i}Y_{j}\phi_{Jk}^{per}(X_{il})\phi_{Js}^{per}(X_{jl})}{\hat{h}_{n}(\textbf{X}_{i})\hat{h}_{n}(\textbf{X}_{j})}\right]\mathop{\rightarrow}\limits_{n\rightarrow\infty}
\mathbb{E}\left[\frac{Y\phi_{Jk}^{per}(X_{l})}{h(\textbf{X})}\right]\mathbb{E}\left[\frac{Y\phi_{Js}^{per}(X_{l})}{h(\textbf{X})}\right]\,.
\end{equation}
Similarly, for $i=j$:

\begin{equation}
\nonumber
\mathbb{E}\left[\frac{Y_{i}^{2}\phi_{Jk}^{per}(X_{il})\phi_{Js}^{per}(X_{il})}{\hat{h}_{n}(\textbf{X}_{i})^{2}}\right]\mathop{\rightarrow}\limits_{n\rightarrow\infty}
\mathbb{E}\left[\frac{Y^{2}\phi_{Jk}^{per}(X_{l})\phi_{Js}^{per}(X_{l})}{h(\textbf{X})^{2}}\right]\,.
\end{equation}
Therefore, it follows that:

\begin{equation}
\nonumber
Cov\left(\hat{c}_{Jk}^{(l)}\,,\,\hat{c}_{Js}^{(l)} \right) \mathop{\rightarrow}\limits_{n\rightarrow\infty} 0\,,
\end{equation}
as desired.
\newpage

\section{Proof of Proposition 5.} \label{proof:Prop5}
Let's assume conditions (\textbf{A1})-(\textbf{A5}) and (\textbf{Ak1})-(\textbf{Ak4}) are satisfied. For $i=1,...,n$, define:

\begin{eqnarray}
  K_{J}(x,y) &=& 2^{J}\sum_{k}\phi(2^{J}x-k)\phi(2^{J}y-k) \\
  Z_{i}(\textbf{x}) &=& \frac{y_{i}}{\hat{h}_{n}(\textbf{x}_{i})}\left(\sum_{l=1}^{p}K_{J}(X_{il},x_{l})\right)-\mathbb{E}\left[\frac{y_{1}}{\hat{h}_{n}(\textbf{x}_{1})}\left(\sum_{l=1}^{p}K_{J}(X_{1l},x_{l})\right) \right]\,.
\end{eqnarray}
Since $\textbf{X}_{1},...,\textbf{X}_{n}$ are iid, $Z_{i}(\textbf{x}),\,\,i=1,...,n$ are iid with $\mathbb{E}[Z_{i}(\textbf{x})]=0$. From the definition of $\hat{f}_{J}(\textbf{x})$ and $Z_{i}(\textbf{x})$, after some algebra it is possible to get:

\begin{eqnarray}
\nonumber
\mathbb{E}\left[||\hat{f}_{J}(\textbf{x})-\mathbb{E}[\hat{f}_{J}(\textbf{x})]||_{2}^{2}\right] &\leq& \mathbb{E}\left[\int_{[0,1]^{p}}\left\{\mid (\hat{\beta}_{0}-\mathbb{E}[\hat{\beta}_{0}])\left(1-2^{-\frac{J}{2}}
\sum_{k=0}^{2^{J}-1}\sum_{l=1}^{p}\phi_{Jk}^{per}(x_{l})\right)\mid + \mid\frac{1}{n}\sum_{i=1}^{n}Z_{i}(\textbf{x})\mid\right\}^{2}d\textbf{x}\right]\,, \\
\nonumber
&\leq& 2\mathbb{E}\left[(\hat{\beta}_{0}-\mathbb{E}[\hat{\beta}_{0}])^{2}\right]\int_{[0,1]^{p}}\left(1-2^{-\frac{J}{2}}\sum_{k=0}^{2^{J}-1}\sum_{l=1}^{p}\phi_{Jk}^{per}(x_{l})\right)^{2}d\textbf{x} \\
& & +\frac{2}{n^{2}}\int_{[0,1]^{p}}\mathbb{E}\left[\mid  \sum_{i=1}^{n}Z_{i}(\textbf{x})\mid^{2}\right]d\textbf{x}\,. \label{eq:ProofProp5.1}
\end{eqnarray}
Denote:
\begin{eqnarray}
\nonumber 
  S_{f1} &=& \int_{[0,1]^{p}}\left(1-2^{-\frac{J}{2}}\sum_{k=0}^{2^{J}-1}\sum_{l=1}^{p}\phi_{Jk}^{per}(x_{l})\right)^{2}d\textbf{x}\,, \\
\nonumber
  S_{f2} &=& \mathbb{E}\left[(\hat{\beta}_{0}-\mathbb{E}[\hat{\beta}_{0}])^{2}\right] = Var\left(\hat{\beta}_{0}\right)\,,\\
\nonumber
  S_{f3} &=& \frac{2}{n^{2}}\int_{[0,1]^{p}}\mathbb{E}\left[\mid  \sum_{i=1}^{n}Z_{i}(\textbf{x})\mid^{2}\right]\,.
\end{eqnarray}
\subsection*{Computations for $S_{f1}$}

Expanding the squared argument for $S_{f1}$, it follows:

\begin{eqnarray}
\nonumber 
   S_{f1} &=& \int_{[0,1]^{p}}\left(1-2^{1-\frac{J}{2}}\sum_{l=1}^{p}\sum_{k=0}^{2^{J}-1} \phi_{Jk}^{per}(x_{l}) + \sum_{l=1}^{p}\sum_{k_{1}=0}^{2^{J}-1}\sum_{m=1}^{p}\sum_{k_{2}=0}^{2^{J}-1}\phi_{Jk_{1}}^{per}(x_{l})\phi_{Jk_{2}}^{per}(x_{m})\right)d\textbf{x}\,, \\
\nonumber
   &=& 1-2^{1-\frac{J}{2}}\sum_{l=1}^{p}\sum_{k=0}^{2^{J}-1}\int_{0}^{1}\phi_{Jk}^{per}(x_{l})dx_{l}+ \sum_{l=1}^{p}\sum_{k_{1}=0}^{2^{J}-1}\sum_{m=1}^{p}\sum_{k_{2}=0}^{2^{J}-1}\int_{0}^{1}\int_{0}^{1}\phi_{Jk_{1}}^{per}(x_{l})\phi_{Jk_{2}}^{per}(x_{m})dx_{l}dx_{m}\,.
\end{eqnarray}
Since $\int_{0}^{1}\mid \phi_{Jk}^{per}(x_{l})\mid dx_{l}\leq C_{\phi}2^{-\frac{J}{2}}$ and $\left\{ \phi^{per}_{J,k}(x), k =0,...,2^{J}-1\right\}$ are orthonormal, it follows:

\begin{equation}\label{eq:ProofProp5.2}
S_{f1} = (p-1)^{2}+p^{2}\left(2^{J}-1\right) = \mathcal{O}\left(2^{J}\right)\,.
\end{equation}

\subsection*{Computations for $S_{f2}$}

Using the identity $Var(X)=\mathbb{E}[X^{2}]-(\mathbb{E}[X])^{2}$, since $\hat{\beta}_{0}=\frac{1}{n}\sum_{i=1}^{n}\frac{y_{i}}{\hat{h}_{n}(\textbf{x}_{i})}$ it is possible to show:

\begin{eqnarray}
\nonumber
\mathbb{E}\left[\hat{\beta}_{0}^{2}\right] &=& \frac{1}{n^{2}}\sum_{i=1}^{n}\sum_{j=1}^{n}\frac{Y_{i}Y_{j}}{\hat{h}_{n}(\textbf{X}_{i})\hat{h}_{n}(\textbf{X}_{j})}\,, \\
\nonumber
& \leq & \frac{\left(|\beta_{0}|+pM_{f}\right)^{2}+\sigma^{2}}{n^{2}}\sum_{i=1}^{n}\mathbb{E}\left[\frac{1}{\hat{h}_{n}(\textbf{X}_{i})^{2}}\right]+ \frac{2}{n^{2}}\sum_{i<j}^{n}\mathbb{E}\left[\frac{Y_{i}Y_{j}}{\hat{h}_{n}(\textbf{X}_{i})\hat{h}_{n}(\textbf{X}_{j})}\right]\,, \\
\nonumber
& \leq & \frac{\left(|\beta_{0}|+pM_{f}\right)^{2}+\sigma^{2}}{n}\mathbb{E}\left[\frac{1}{\hat{h}_{n}(\textbf{X})^{2}}\right]+ \frac{2}{n^{2}}\sum_{i<j}^{n}\mathbb{E}\left[\frac{Y_{i}Y_{j}}{\hat{h}_{n}(\textbf{X}_{i})\hat{h}_{n}(\textbf{X}_{j})}\right]\,, \\
\nonumber
& \leq & \frac{\left(|\beta_{0}|+pM_{f}\right)^{2}+\sigma^{2}}{n}\mathbb{E}\left[\frac{1}{\hat{h}_{n}(\textbf{X})^{2}}-\frac{1}{h(\textbf{X})^{2}}\right]+ \frac{\left(|\beta_{0}|+pM_{f}\right)^{2}+\sigma^{2}}{n}\mathbb{E}\left[\frac{1}{h(\textbf{X})^{2}}\right] \\
\nonumber
& & + \frac{2}{n^{2}}\sum_{i<j}^{n}\mathbb{E}\left[\frac{Y_{i}Y_{j}}{\hat{h}_{n}(\textbf{X}_{i})\hat{h}_{n}(\textbf{X}_{j})}\right]\,.
\end{eqnarray}
Now, since $|\mathbb{E}\left[\frac{1}{\hat{h}_{n}(\textbf{X})^{2}}-\frac{1}{h(\textbf{X})^{2}} \right]|\leq Cn^{-\frac{\beta}{2\beta+p}}$ and $h(\textbf{x})>\epsilon_{h}$, it follows:

\begin{equation}
\nonumber
\mathbb{E}\left[\hat{\beta}_{0}^{2}\right]  \leq C_{1}n^{-\frac{3\beta+p}{2\beta+p}} + C_{2}n^{-1}+\frac{2}{n^{2}}\sum_{i<j}^{n}\mathbb{E}\left[\frac{Y_{i}Y_{j}}{\hat{h}_{n}(\textbf{X}_{i})\hat{h}_{n}(\textbf{X}_{j})}\right]\,,
\end{equation}
for $C_{1}=C\cdot \left(|\beta_{0}|+pM_{f}\right)^{2}+\sigma^{2}$ and $C_{2}= \frac{\left(|\beta_{0}|+pM_{f}\right)^{2}+\sigma^{2}}{\epsilon_{h}^{2}}$.

\medskip

Since $n\delta^{p}$ uniformly converges to $\infty$, $\hat{h}_{n}(\textbf{X}_{i}) \approx \hat{h}_{n-1}^{(-i)}(\textbf{X}_{i})$, for $n$ large. The notation $\approx$ means that the ratio between the lhs and the rhs terms goes to 1 as $n \rightarrow \infty$. Also, since we have an iid sample, it holds:

\begin{eqnarray}
\nonumber
\mathbb{E}\left[\frac{Y_{i}Y_{j}}{\hat{h}_{n}(\textbf{X}_{i})\hat{h}_{n}(\textbf{X}_{j})}\right] &=& \mathbb{E}_{\textbf{X}_{i},\textbf{X}_{j}}\left[\mathbb{E}_{\textbf{X}^{(-i,-j)}|\textbf{X}_{i},\textbf{X}_{j}}\left(\frac{Y_{i}Y_{j}}{\hat{h}_{n-1}^{(-i)}(\textbf{X}_{i})\hat{h}_{n-1}^{(-j)}(\textbf{X}_{j})}|\textbf{X}_{i},\textbf{X}_{j} \right)\right]\,, \\
\nonumber
&=&  \mathbb{E}_{\textbf{X}_{i},\textbf{X}_{j}}\left[\mathbb{E}_{\textbf{X}^{(-i,-j)}}\left(\frac{Y_{i}}{\hat{h}_{n-1}^{(-i)}(\textbf{X}_{i})}|\textbf{X}_{i},\textbf{X}_{j} \right)
\mathbb{E}_{\textbf{X}^{(-i,-j)}}\left(\frac{Y_{j}}{\hat{h}_{n-1}^{(-j)}(\textbf{X}_{j})}|\textbf{X}_{i},\textbf{X}_{j} \right)\right]\,, \\
\nonumber
&\approx& \mathbb{E}_{\textbf{X}_{i},\textbf{X}_{j}}\left[\mathbb{E}_{\textbf{X}^{(-i,-j)}}\left(\frac{Y_{i}}{\hat{h}_{n}(\textbf{X}_{i})}\right)
\mathbb{E}_{\textbf{X}^{(-i,-j)}}\left(\frac{Y_{j}}{\hat{h}_{n}(\textbf{X}_{j})}\right)\right]\,, \\
\nonumber
&\approx& \left(\mathbb{E}\left[\frac{Y}{\hat{h}_{n}(\textbf{X})} \right] \right)^{2}\,.
\end{eqnarray}

This implies:

\begin{eqnarray}
  \mathbb{E}\left[\hat{\beta}_{0}^{2}\right] &\leq & C^{*}\left(n^{-\frac{3\beta+p}{2\beta+p}} + n^{-1}\right) + \frac{n(n-1)}{n^{2}}\left(\mathbb{E}\left[\frac{Y}{\hat{h}_{n}(\textbf{X})} \right] \right)^{2}\,, \label{eq:ProofProp5.3}
\end{eqnarray}
for some $C^{*}> \max\left\{C_{1},C_{2}\right\}>0$. Similarly, it follows:

\begin{eqnarray}
\nonumber
\mathbb{E}\left[\hat{\beta}_{0}\right] &=& \mathbb{E}\left[\frac{\beta_{0}+\sum_{l=1}^{p}f_{l}(x_{l})+\epsilon}{\hat{h}_{n}(\textbf{X})}\right]\,, \\
\nonumber
& = & \mathbb{E}\left[\frac{Y}{\hat{h}_{n}(\textbf{X})}\right]\,.
\end{eqnarray}
The last result, together with (\ref{eq:ProofProp5.3}) imply:

\begin{eqnarray}
\nonumber
\mathbb{E}\left[\hat{\beta}_{0}^{2}\right]-\left(\mathbb{E}\left[\hat{\beta}_{0}\right]\right)^{2} &\leq& C^{*}\left(n^{-\frac{3\beta+p}{2\beta+p}} + n^{-1}\right) + \frac{n(n-1)}{n^{2}}\left(\mathbb{E}\left[\frac{Y}{\hat{h}_{n}(\textbf{X})} \right] \right)^{2}-\left(\mathbb{E}\left[\frac{Y}{\hat{h}_{n}(\textbf{X})}\right]\right)^{2}\,, \\
\nonumber
&\leq& C^{*}\left(n^{-\frac{3\beta+p}{2\beta+p}} + n^{-1}\right)-\frac{1}{n}\left(\mathbb{E}\left[\frac{Y}{\hat{h}_{n}(\textbf{X})}\right]\right)^{2}\,, \\
\nonumber
&\leq& C^{*}\left(n^{-\frac{3\beta+p}{2\beta+p}} + n^{-1}\right)\,, \\
S_{f2} &=& \mathcal{O}\left(n^{-1}\right)\,. \label{eq:ProofProp5.4}
\end{eqnarray}
Thus, from (\ref{eq:ProofProp5.2}) and (\ref{eq:ProofProp5.4}), it follows that:

\begin{equation}
\nonumber
S_{f1}S_{f2}=\mathcal{O}\left(2^{J}n^{-1}\right)\,.
\end{equation}

\subsection*{Computations for $S_{f3}$}

From Rosenthal's inequality, $\exists\, C(2)>0$ such that:

\begin{eqnarray}
\nonumber
\frac{2}{n^{2}}\int_{[0,1]^{p}}\mathbb{E}\left[\left|\sum_{i=1}^{n}Z_{i}(\textbf{x})\right|^{2}\right] &\leq& \frac{4C(2)}{n^{2}}\int_{[0,1]^{p}}\sum_{i=1}^{n}\mathbb{E}\left[Z_{i}(\textbf{x})^{2} \right]d\textbf{x}\,, \\
\nonumber
&\leq&  \frac{4C(2)}{n^{2}}\sum_{i=1}^{n}\int_{[0,1]^{p}}\mathbb{E}\left[Z_{i}(\textbf{x})^{2} \right]d\textbf{x}\,.
\end{eqnarray}
By the definition of $Z_{i}(\textbf{x})$, it follows:

\begin{eqnarray}
\nonumber
\int_{[0,1]^{p}}\mathbb{E}\left[Z_{i}(\textbf{x})^{2} \right]d\textbf{x} &\leq & \sum_{l=1}^{p}\sum_{k_{1}=0}^{2^{J}-1}\sum_{m=1}^{p}\sum_{k_{2}=0}^{2^{J}-1}\mathbb{E}\left[\frac{Y_{i}^{2}\phi_{Jk_{1}}^{per}(X_{il})\phi_{Jk_{2}}^{per}(X_{im})}{\hat{h}_{n}(\textbf{X}_{i})^{2}} \right]\int_{[0,1]^{p}}\phi_{Jk_{1}}^{per}(x_{l})\phi_{Jk_{2}}^{per}(x_{m})d\textbf{x}\,.
\end{eqnarray}
From the orthonormality of the scaling functions $\left\{ \phi^{per}_{J,k}(x), k =0,...,2^{J}-1\right\}$ and (\ref{eq:proof1}), it follows:
\begin{equation}
\nonumber
\int_{[0,1]^{p}}\phi_{Jk_{1}}^{per}(x_{l})\phi_{Jk_{2}}^{per}(x_{m})d\textbf{x} =
        \begin{cases}
            1 & k_{1}=k_{2}  \quad l=m \\
            0 & k_{1}\neq k_{2}  \quad l=m \\
            2^{-J} & k_{1}=k_{2}  \quad l\neq m \\
            2^{-J} & k_{1}\neq k_{2}  \quad l\neq m
        \end{cases}
\end{equation}
Therefore,

\begin{eqnarray}
\nonumber
\int_{[0,1]^{p}}\mathbb{E}\left[Z_{i}(\textbf{x})^{2} \right]d\textbf{x} &\leq& \sum_{l=1}^{p}\sum_{k=0}^{2^{J}-1}\left(\mathbb{E}\left[\frac{Y_{i}^{2}\phi_{Jk}^{per}(X_{il})^{2}}{\hat{h}_{n}(\textbf{X}_{i})^{2}}\right]\right) \\
\nonumber
& & + 2^{-J}\sum_{l\neq m}^{p}\sum_{k=0}^{2^{J}-1}\left(\mathbb{E}\left[\frac{Y_{i}^{2}\phi_{Jk}^{per}(X_{il})\phi_{Jk}^{per}(X_{im})}{\hat{h}_{n}(\textbf{X}_{i})^{2}}\right]\right) \\
\nonumber
& & + 2^{-J}\sum_{l\neq m}^{p}\sum_{k_{1}\neq k_{2}}^{2^{J}-1}\left(\mathbb{E}\left[\frac{Y_{i}^{2}\phi_{Jk_{1}}^{per}(X_{il})\phi_{Jk_{2}}^{per}(X_{im})}{\hat{h}_{n}(\textbf{X}_{i})^{2}} \right] \right)\,.
\end{eqnarray}
Since $\mathop{\sup}\limits_{\textbf{x}\in [0,1]^{p}}\left\{\beta_{0}+\sum_{l=1}^{p}f_{l}(x_{l})\right\}\leq\left(|\beta_{0}|+pM_{f}\right)$, we can show:

\begin{eqnarray}
\nonumber
\mathbb{E}\left[\frac{Y_{i}^{2}\phi_{Jk}^{per}(X_{il})^{2}}{\hat{h}_{n}(\textbf{X}_{i})^{2}}\right] &\leq& \left(\left(|\beta_{0}|+pM_{f}\right)^{2}+\sigma^{2}\right)\mathbb{E}\left[ \frac{\phi_{Jk}^{per}(X_{il})^{2}}{\hat{h}_{n}(\textbf{X}_{i})^{2}}\right]\,, \\
\nonumber
&\leq& C_{1}\mathbb{E}\left[\phi_{Jk}^{per}(X_{il})^{2}\left(\frac{1}{\hat{h}_{n}(\textbf{X}_{i})^{2}}-\frac{1}{h(\textbf{X})^{2}}\right)\right]\,, \\
\nonumber
&\leq& C_{1}\cdot C\cdot n^{-\frac{\beta}{2\beta+p}}\mathbb{E}\left[\phi_{Jk}^{per}(X_{il})^{2}\right]\,, \\
&\leq& C_{1}\cdot C\cdot Mn^{-\frac{\beta}{2\beta+p}}
\end{eqnarray}
for $C_{1}=\left(\left(|\beta_{0}|+pM_{f}\right)^{2}+\sigma^{2}\right)$ and $M$ as the upper bound of the density $h(\textbf{x})$ from assumption (\textbf{A5}). Similarly, when $l\neq m$, it follows:

\begin{eqnarray}
\nonumber
\mathbb{E}\left[\frac{Y_{i}^{2}\phi_{Jk}^{per}(X_{il})\phi_{Jk}^{per}(X_{im})}{\hat{h}_{n}(\textbf{X}_{i})^{2}}\right] &\leq& \left(\left(|\beta_{0}|+pM_{f}\right)^{2}+\sigma^{2}\right)
\mathbb{E}\left[ \frac{\phi_{Jk}^{per}(X_{il})\phi_{Jk}^{per}(X_{im})}{\hat{h}_{n}(\textbf{X}_{i})^{2}}\right]\,, \\
\nonumber
 &\leq& C_{1}\mathbb{E}\left[ \phi_{Jk}^{per}(X_{il})\phi_{Jk}^{per}(X_{im})\left(\frac{1}{\hat{h}_{n}(\textbf{X}_{i})^{2}}-\frac{1}{h(\textbf{X})^{2}}\right)+
 \frac{\phi_{Jk}^{per}(X_{il})\phi_{Jk}^{per}(X_{im})}{h(\textbf{X})^{2}}\right]\,, \\
\nonumber
&\leq& C_{1}\cdot C\cdot n^{-\frac{\beta}{2\beta+p}}\mathbb{E}\left[ \phi_{Jk}^{per}(X_{il})\phi_{Jk}^{per}(X_{im})\right]+\frac{C_{1}}{\epsilon_{h}^{2}}\mathbb{E}\left[ \phi_{Jk}^{per}(X_{il})\phi_{Jk}^{per}(X_{im})\right]\,, \\
\nonumber
&\leq& C_{1}\cdot C\cdot M2^{-J}n^{-\frac{\beta}{2\beta+p}}+\frac{C_{1}}{\epsilon_{h}^{2}}M2^{-J}\,.
\end{eqnarray}
In the case $k_{1}\neq k_{2}  \quad l\neq m$, it is possible to show:

\begin{eqnarray}
\nonumber
\mathbb{E}\left[\frac{Y_{i}^{2}\phi_{Jk_{1}}^{per}(X_{il})\phi_{Jk_{2}}^{per}(X_{im})}{\hat{h}_{n}(\textbf{X}_{i})^{2}} \right] &\leq& C_{1}\cdot
\mathbb{E}\left[ \frac{\phi_{Jk_{1}}^{per}(X_{il})\phi_{Jk_{2}}^{per}(X_{im})}{\hat{h}_{n}(\textbf{X}_{i})^{2}}\right]\,, \\
\nonumber
&\leq& C_{1}\mathbb{E}\left[ \phi_{Jk_{1}}^{per}(X_{il})\phi_{Jk_{2}}^{per}(X_{im})\left(\frac{1}{\hat{h}_{n}(\textbf{X}_{i})^{2}}-\frac{1}{h(\textbf{X})^{2}}\right)+
 \frac{\phi_{Jk_{1}}^{per}(X_{il})\phi_{Jk_{2}}^{per}(X_{im})}{h(\textbf{X})^{2}}\right]\,, \\
\nonumber
&\leq& C_{1}\cdot C\cdot n^{-\frac{\beta}{2\beta+p}}\mathbb{E}\left[ \phi_{Jk_{1}}^{per}(X_{il})\phi_{Jk_{2}}^{per}(X_{im})\right]+\frac{C_{1}}{\epsilon_{h}^{2}}\mathbb{E}\left[ \phi_{Jk_{1}}^{per}(X_{il})\phi_{Jk_{2}}^{per}(X_{im})\right]\,, \\
\nonumber
&\leq& C_{1}\cdot C\cdot M2^{-J}n^{-\frac{\beta}{2\beta+p}}+\frac{C_{1}}{\epsilon_{h}^{2}}M2^{-J}\,.
\end{eqnarray}
The last set of results imply:

\begin{eqnarray}
\nonumber
\int_{[0,1]^{p}}\mathbb{E}\left[Z_{i}(\textbf{x})^{2} \right]d\textbf{x} &\leq& p\cdot2^{J}\cdot C_{1}\cdot C\cdot Mn^{-\frac{\beta}{2\beta+p}} \\
\nonumber
& & +p(p-1)\left\{C_{1}\cdot C\cdot M2^{-J}n^{-\frac{\beta}{2\beta+p}}+\frac{C_{1}}{\epsilon_{h}^{2}}M2^{-J} \right\} \\
\nonumber
& & +p(p-1)(2^{J}-1)\left\{C_{1}\cdot C\cdot M2^{-J}n^{-\frac{\beta}{2\beta+p}}+\frac{C_{1}}{\epsilon_{h}^{2}}M2^{-J} \right\}\,, \\
\nonumber
&\leq& p\cdot2^{J}\cdot C_{1}\cdot C\cdot Mn^{-\frac{\beta}{2\beta+p}}+p(p-1)\left\{C_{1}\cdot C\cdot M\cdot n^{-\frac{\beta}{2\beta+p}}+\frac{C_{1}}{\epsilon_{h}^{2}}M\right\}\,, \\
\nonumber
&\leq& C^{*}\left(2^{J}n^{-\frac{\beta}{2\beta+p}}+n^{-\frac{\beta}{2\beta+p}}+1\right)\,,
\end{eqnarray}
for $C^{*}=\max\left\{p\,C_{1}\,C\,M\,,\, p(p-1)\,C_{1}\,C\,M\,,\,p(p-1)\frac{C_{1}}{\epsilon_{h}^{2}}M\right\}>0$. Finally, we obtain:

\begin{eqnarray}
\nonumber
\frac{4C(2)}{n^{2}}\sum_{i=1}^{n}\int_{[0,1]^{p}}\mathbb{E}\left[Z_{i}(\textbf{x})^{2} \right]d\textbf{x} &\leq& \frac{4C(2)}{n}C^{*}\left(2^{J}n^{-\frac{\beta}{2\beta+p}}+n^{-\frac{\beta}{2\beta+p}}+1\right)\,, \\
\nonumber
&\leq&  C^{**}\left(2^{J}n^{-\frac{3\beta+p}{2\beta+p}}+n^{-\frac{3\beta+p}{2\beta+p}}+n^{-1}\right)\,, \\
S_{f3} &=& \mathcal{O}\left(2^{J}n^{-\frac{3\beta+p}{2\beta+p}}+n^{-\frac{3\beta+p}{2\beta+p}}+n^{-1}\right)\,, \label{eq:ProofProp5.5}
\end{eqnarray}
for $C^{**}=4C(2)\,C^{*}\,>0 $.

\medskip

Finally, from (\ref{eq:ProofProp5.1}),(\ref{eq:ProofProp5.3}) and (\ref{eq:ProofProp5.5}), it follows:

\begin{eqnarray}
\nonumber 
   \mathbb{E}\left[||\hat{f}_{J}(\textbf{x})-\mathbb{E}[\hat{f}_{J}(\textbf{x})]||_{2}^{2}\right] &\leq& \mathcal{O}\left(2^{J}n^{-1}\right)+ \mathcal{O}\left(2^{J}n^{-\frac{3\beta+p}{2\beta+p}}+n^{-1}\right) \\
   &\leq&  \mathcal{O}\left(2^{J}n^{-1}\right)\,. \label{eq:ProofProp5.6}
\end{eqnarray}
which completes the proof.

\newpage

\section{Proof of Proposition 6.} \label{proof:Prop6}

Suppose that in addition to assumptions (\textbf{A1})-(\textbf{A5}) and (\textbf{Ak1})-(\textbf{Ak4}), the following conditions are satisfied:

\begin{enumerate}
  \item $\exists\,\Phi$, bounded and non-increasing function in $\mathbb{R}$ such that $\int\Phi(|u|)du<\infty$ and $|\phi(u)|\leq \Phi(|u|)$ almost everywhere (a.e.).
  \item In addition, $\int_{\mathbb{R}}|u|^{N+1}\Phi(|u|)du<\infty$ for some $N\geq0$.
  \item $\exists\,F$, integrable, such that $|K(x,y)|\leq F(x-y)$, $\forall x,y \in \mathbb{R}$.
  \item Suppose $\phi$ satisfies:
  \begin{enumerate}
    \item $\sum_{k}|\hat{\phi}(\xi+2k\pi)|^{2}=1$, a.e., where $\hat{\phi}$ denotes the Fourier transform of the scaling function $\phi$. \label{ass:w1}
    \item $\hat{\phi}(\xi)=\hat{\phi}(\frac{\xi}{2})m_{0}(\frac{\xi}{2})$, where $m_{0}(\xi)$ is a $2\pi$-periodic function and $m_{0}\,\in\,\mathbb{L}_{2}(0,2\pi)$.\label{ass:w2}
  \end{enumerate}
  \item $\int_{\mathbb{R}}x^{k}\psi(x)dx=0$, for $k=0,1,...,N$, $N\geq1$ where $\psi$ is the mother wavelet corresponding to $\phi$.
  \item The functions $\left\{f_{l}\right\}_{l=1}^{p}$, are such that $f_{l}\,\in\,L_{\infty}([0,1])$ and $f_{l}\,\in\,W_{\infty}^{m+1}([0,1])\,,\,m \geq N$, where $W_{\infty}^{m}([0,1])$ denotes the space of functions that are $m$-times weakly-differentiable and $f_{l}^{(k)}\,\in\,L_{\infty}([0,1])\,,\, k=1,...,m$.
  \item $\theta_{\phi}(x):=\sum_{k}|\phi(x-k)|$ such that $||\theta_{\phi}||_{\infty}<\infty$.
\end{enumerate}
Then under Corollary 8.2 \cite{Hardle1998}, if $f\in W_{\infty}^{N+1}([0,1])$ then $||K_{J}f-f||_{\infty}^{p}=\mathcal{O}\left(2^{-pJ(N+1)}\right)\,,\,p\geq1$. This implies:

\medskip

\begin{equation}\label{eq:proofProp6.1}
||\mathbb{E}[\hat{f}_{J}(\textbf{x})]-f(\textbf{x}) ||_{2}^{2} = \mathcal{O}\left(2^{2J}n^{-\frac{2\beta}{2\beta+p}}+2^{-2J(N+1)}+n^{-\frac{\beta}{2\beta+p}}2^{-J(N+1)}\right)\,,
\end{equation}
\medskip
for $f(\textbf{x})=\beta_{0}+\sum_{l=1}^{p}f_{l}(x_{l})$.

\subsection*{Proof}

Define $f_{lJ}(x_{l}):=K_{J}f_{l}(x_{l})=\int_{0}^{1}f_{l}(u)K_{J}\left(x_{l},u\right)du$. Suppose a fixed $\textbf{x}$, then:

\begin{eqnarray}
\nonumber
\mathbb{E}[\hat{f}_{J}(\textbf{x})]-f(\textbf{x}) &=& bias\left(\hat{\beta}_{0}\right)+\sum_{l=1}^{p}\sum_{k=0}^{2^{J}-1}bias\left(\hat{c}_{Jk}^{(l)}\right)\phi_{Jk}^{per}(x_{l})+
\sum_{l=1}^{p}\left(f_{lJ}(x_{l})-f_{l}(x_{l})\right)\,.
\end{eqnarray}
Furthermore, since $\mathbb{E}\left[\frac{\sum_{l=1}^{p}f_{l}(X_{l})}{h(\textbf{X})}\right]=0$, it follows:

\begin{eqnarray}
\nonumber
bias\left(\hat{\beta}_{0}\right) &\leq& |\beta_{0}|Cn^{-\frac{\beta}{2\beta+p}}+\mathbb{E}_{\textbf{X}}\left[\sum_{l=1}^{p}f_{l}(x_{l})\mathbb{E}_{\textbf{X}_{1},...,\textbf{X}_{n}}\left( \frac{1}{\hat{h}_{n}(\textbf{X})}-\frac{1}{h(\textbf{X})}\right) \right] \,,\\
\nonumber
&\leq& \left(|\beta_{0}|+pM_{f}\right)Cn^{-\frac{\beta}{2\beta+p}}\,.
\end{eqnarray}
Similarly, following the same argument for $bias\left(\hat{c}_{Jk}^{(l)}\right)$, it is possible to show:

\begin{eqnarray}
\nonumber
bias\left(\hat{c}_{Jk}^{(l)}\right) &\leq & \left(|\beta_{0}|+pM_{f}\right)C2^{-\frac{J}{2}}n^{-\frac{\beta}{2\beta+p}}\,.
\end{eqnarray}
Therefore, this implies:

\begin{eqnarray}
\nonumber
\mathbb{E}[\hat{f}_{J}(\textbf{x})]-f(\textbf{x}) &\leq& C_{1}^{*}n^{-\frac{\beta}{2\beta+p}}+C_{1}^{*}2^{-\frac{J}{2}}n^{-\frac{\beta}{2\beta+p}}\sum_{l=1}^{p}\sum_{k=0}^{2^{J}-1}\left|\phi_{Jk}^{per}(x_{l})\right|+
\sum_{l=1}^{p}\left|K_{J}f_{l}(x_{l})-f_{l}(x_{l}) \right|\,, \\
\nonumber
\left(\mathbb{E}[\hat{f}_{J}(\textbf{x})]-f(\textbf{x})\right)^{2} &\leq& C_{1}^{**}n^{-\frac{2\beta}{2\beta+p}}+C_{1}^{**}2^{-J}n^{-\frac{2\beta}{2\beta+p}}\sum_{l=1}^{p}\sum_{k_{1}=0}^{2^{J}-1}\sum_{m=1}^{p}\sum_{k_{2}=0}^{2^{J}-1}
\left|\phi_{Jk_{1}}^{per}(x_{l})\right|\left|\phi_{Jk_{2}}^{per}(x_{m})\right| \\
\nonumber
& & + \left(\sum_{l=1}^{p}\left|K_{J}f_{l}(x_{l})-f_{l}(x_{l})\right|\right)^{2}
+ 2C_{1}^{*}2^{-\frac{J}{2}}n^{-\frac{2\beta}{2\beta+p}}\sum_{l=1}^{p}\sum_{k=0}^{2^{J}-1}\left|\phi_{Jk}^{per}(x_{l})\right| \\
\nonumber
& & + 2C_{1}^{*}n^{-\frac{\beta}{2\beta+p}}\sum_{l=1}^{p}\left|K_{J}f_{l}(x_{l})-f_{l}(x_{l}) \right| \\
\nonumber
& & + 2C_{1}^{*}2^{-\frac{J}{2}}n^{-\frac{\beta}{2\beta+p}}\sum_{l=1}^{p}\sum_{k=0}^{2^{J}-1}\sum_{m=1}^{p}\left|\phi_{Jk}^{per}(x_{l})\right|\left|K_{J}f_{m}(x_{m})-f_{m}(x_{m})\right|\,,
\end{eqnarray}
where $C_{1}^{*}=\left(|\beta_{0}|+pM_{f}\right)C$ and $C_{1}^{**}=\left(|\beta_{0}|+pM_{f}\right)^{2}C^{2}$ are positive constants independent of $J$ and $n$. Furthermore, since $\int_{0}^{1}\left|\phi_{Jk}^{per}(u)\right|du=2^{-\frac{J}{2}}C_{\phi}$, $C_{\phi}$, it follows:

\begin{equation}
\nonumber
\int_{[0,1]^{p}}|\phi_{Jk_{1}}^{per}(x_{l})||\phi_{Jk_{2}}^{per}(x_{m})|d\textbf{x} =
        \begin{cases}
            1 & k_{1}=k_{2}  \quad l=m \\
            2^{J}||\theta_{\phi}||_{\infty}^{2} & k_{1}\neq k_{2}  \quad l=m \\
            2^{-J}C_{\phi}^{2} & k_{1}=k_{2}  \quad l\neq m \\
            2^{-J}C_{\phi}^{2} & k_{1}\neq k_{2}  \quad l\neq m
        \end{cases}
\end{equation}
Using the last set of equations, we obtain:

\begin{eqnarray}
\nonumber
\int_{[0,1]^{p}}\left(\mathbb{E}[\hat{f}_{J}(\textbf{x})]-f(\textbf{x})\right)^{2}d\textbf{x} &\leq& C_{1}^{**}n^{-\frac{2\beta}{2\beta+p}} + pC_{1}^{**}n^{-\frac{2\beta}{2\beta+p}} + pC_{1}^{**}n^{-\frac{2\beta}{2\beta+p}}2^{J}(2^{J}-1)||\theta_{\phi}||_{\infty}^{2} \\
\nonumber
& & + C_{\phi}^{2}C_{1}^{**}n^{-\frac{2\beta}{2\beta+p}}p(p-1) + \left\|\sum_{l=1}^{p}\left|K_{J}f_{l}(x_{l})-f_{l}(x_{l})\right| \right\|_{2}^{2} \\
\nonumber
& & + 2pC_{\phi}C_{1}^{*}n^{-\frac{2\beta}{2\beta+p}} + 2C_{1}^{*}n^{-\frac{\beta}{2\beta+p}}\left\|\sum_{l=1}^{p}\left|K_{J}f_{l}(x_{l})-f_{l}(x_{l})\right| \right\|_{1} \\
\nonumber
& & + 2C_{1}^{*}2^{-\frac{J}{2}}n^{-\frac{\beta}{2\beta+p}}\sum_{l=1}^{p}\sum_{k=0}^{2^{J}-1}\sum_{m=1}^{p}\int_{[0,1]^{p}}\left|\phi_{Jk}^{per}(x_{l})\right|\left|K_{J}f_{m}(x_{m})-f_{m}(x_{m})\right|d\textbf{x}\,.
\end{eqnarray}
Using the properties of $L_{p}$ norms and Corollary 8.2 \cite{Hardle1998}, it follows:

\begin{eqnarray}
\nonumber
\left\|\sum_{l=1}^{p}\left|K_{J}f_{l}(x_{l})-f_{l}(x_{l})\right| \right\|_{2}^{2} &\leq& 2\sum_{l=1}^{p}\left\|K_{J}f_{l}(x_{l})-f_{l}(x_{l})\right\|_{2}^{2}\leq C_{*}2^{-2J(N+1)} \,,\\
\nonumber
\left\|\sum_{l=1}^{p}\left|K_{J}f_{l}(x_{l})-f_{l}(x_{l})\right| \right\|_{1} &\leq& \sum_{l=1}^{p}\left\|K_{J}f_{l}(x_{l})-f_{l}(x_{l})\right\|_{1}\leq C_{**}2^{-J(N+1)}\,.
\end{eqnarray}
Therefore, this implies:

\begin{eqnarray}
\nonumber
\int_{[0,1]^{p}}\left(\mathbb{E}[\hat{f}_{J}(\textbf{x})]-f(\textbf{x})\right)^{2}d\textbf{x} &\leq& C_{1}^{**}n^{-\frac{2\beta}{2\beta+p}} + pC_{1}^{**}n^{-\frac{2\beta}{2\beta+p}} + pC_{1}^{**}n^{-\frac{2\beta}{2\beta+p}}2^{J}(2^{J}-1)||\theta_{\phi}||_{\infty}^{2} \\
\nonumber
& & + C_{\phi}^{2}C_{1}^{**}n^{-\frac{2\beta}{2\beta+p}}p(p-1) + C_{*}2^{-2J(N+1)} + 2pC_{\phi}C_{1}^{**}n^{-\frac{2\beta}{2\beta+p}} \\
\nonumber
& & + 2C_{1}^{*}n^{-\frac{\beta}{2\beta+p}}C_{**}2^{-J(N+1)} \\
\nonumber
& & + 2C_{1}^{*}2^{-\frac{J}{2}}n^{-\frac{\beta}{2\beta+p}}\sum_{l=1}^{p}\sum_{k=0}^{2^{J}-1}\int_{[0,1]^{p}}\left|\phi_{Jk}^{per}(x_{l})\right|\left|K_{J}f_{l}(x_{l})-f_{l}(x_{l})\right|d\textbf{x} \\
\nonumber
& & + 2C_{1}^{*}2^{-\frac{J}{2}}n^{-\frac{\beta}{2\beta+p}}\sum_{l\neq m}^{p}\sum_{k=0}^{2^{J}-1}\int_{[0,1]^{p}}\left|\phi_{Jk}^{per}(x_{l})\right|\left|K_{J}f_{m}(x_{m})-f_{m}(x_{m})\right|d\textbf{x}\,, \\
\nonumber
&\leq& C^{***}\left\{n^{-\frac{2\beta}{2\beta+p}}+2^{2J}n^{-\frac{2\beta}{2\beta+p}}+2^{-2J(N+1)}+ n^{-\frac{2\beta}{2\beta+p}}2^{-J(N+1)}\right\} \\
\nonumber
& & + C^{***}\left\{2^{-\frac{J}{2}}n^{-\frac{\beta}{2\beta+p}}\sum_{l=1}^{p}\sum_{k=0}^{2^{J}-1}\int_{[0,1]^{p}}\left|\phi_{Jk}^{per}(x_{l})\right|
\left|K_{J}f_{l}(x_{l})-f_{l}(x_{l})\right|d\textbf{x}\right\} \\
\nonumber
& & + C^{***}\left\{2^{-\frac{J}{2}}n^{-\frac{\beta}{2\beta+p}}\sum_{l\neq m}^{p}\sum_{k=0}^{2^{J}-1}\int_{[0,1]^{p}}\left|\phi_{Jk}^{per}(x_{l})\right|\left|K_{J}f_{m}(x_{m})-f_{m}(x_{m})\right|d\textbf{x}\right\}\,,
\end{eqnarray}
for $C^{***}=\max\left\{p\,C_{1}^{**}\,,\,p\,C_{1}^{**}||\theta_{\phi}||_{\infty}^{2}\,,\,2pC_{\phi}^{2}C_{1}^{**}\,,\,C_{*}\,,\,2C_{1}^{*}C_{**}\right\}>0$, independent of $J$ and $n$.
\medskip

Assumption 7 and Corollary 8.2 \cite{Hardle1998} imply:

\begin{equation}
\nonumber
\int_{[0,1]^{p}}\left|\phi_{Jk}^{per}(x_{l})\right|
\left|K_{J}f_{m}(x_{m})-f_{m}(x_{m})\right|d\textbf{x} \leq
        \begin{cases}
            C2^{-\frac{J}{2}}||\theta_{\phi}||_{\infty}2^{-J(N+1)} & l=m \\
            C\cdot C_{\phi} 2^{-\frac{J}{2}}2^{-J(N+1)} & l\neq m \\
        \end{cases}
\end{equation}
\medskip

Therefore:

\begin{eqnarray}
\nonumber
\int_{[0,1]^{p}}\left(\mathbb{E}[\hat{f}_{J}(\textbf{x})]-f(\textbf{x})\right)^{2}d\textbf{x} &\leq&  \tilde{C}^{***}\left\{n^{-\frac{2\beta}{2\beta+p}}+2^{2J}n^{-\frac{2\beta}{2\beta+p}}+2^{-2J(N+1)}+ n^{-\frac{2\beta}{2\beta+p}}2^{-J(N+1)}+n^{-\frac{\beta}{2\beta+p}}2^{-J(N+1)}\right\}\,, \\
\nonumber
& \leq & \tilde{C}^{***}\left\{2^{2J}n^{-\frac{2\beta}{2\beta+p}}+2^{-2J(N+1)}+n^{-\frac{\beta}{2\beta+p}}2^{-J(N+1)}\right\}\,,
\end{eqnarray}
for $\tilde{C}^{***}=\max\left\{C^{***}\,,\,C||\theta_{\phi}||_{\infty}\,,\,C\cdot C_{\phi} \right\}>0$. Thus,
\medskip
\begin{equation}\label{eq:proofProp6.2}
\left\|\mathbb{E}[\hat{f}_{J}(\textbf{x})]-f(\textbf{x}) \right\|_{2}^{2} = \mathcal{O}\left(2^{2J}n^{-\frac{2\beta}{2\beta+p}}+2^{-2J(N+1)}+n^{-\frac{\beta}{2\beta+p}}2^{-J(N+1)}\right)\,,
\end{equation}
\medskip
which completes the proof.

\subsection*{Remarks}
Note that assumptions \ref{ass:w1} and \ref{ass:w2} are automatically satisfied by choosing the orthonormal basis $\left\{ \phi^{per}_{J,k}(x), k =0,...,2^{J}-1\right\}$. These are explicitly stated to be consistent with results presented in \cite{Hardle1998} that were used to obtain the estimator approximation properties.

\newpage
\section{Proof of Proposition 7.} \label{proof:Prop7}

Define $\mathcal{F}=\left\{f\,|\,f_{l}\,\in\,L_{2}([0,1]),\,f_{l}\,\in\,W_{2}^{N+1}([0,1]),\,-\infty< m_{l}\leq f_{l}\leq M_{l}<\infty\right\}$ where $f(\textbf{x})=\beta_{0}+\sum_{l=1}^{p}f_{l}(x_{l})$. Suppose assumptions 1-7 from Proposition 6 and conditions (\textbf{A1})-(\textbf{A5}), and (\textbf{Ak1})-(\textbf{Ak4}) are satisfied. Then:

\begin{equation}\label{eq:proofProp7.1}
\mathop{\sup}\limits_{f\in \mathcal{F}}\left(\mathbb{E}\left[||\hat{f}_{J}(\textbf{x})-f(\textbf{x}) ||_{2}^{2}\right] \right)\leq \tilde{C}n^{-\left(\frac{\beta}{2\beta+p}\right)\left(\frac{N+1}{N+3}\right)}\,,
\end{equation}
provided (\ref{eq:Risk2}) and (\ref{eq:Risk3}), for $J=J(n)$ such that $2^{J(n)}\simeq n^{\frac{2\beta}{(2\beta+p)(N+3)}}$.

\subsection*{Proof}

For $C>0$ sufficiently large it follows:

\begin{equation}
\nonumber
\mathbb{E}||\hat{f}_{J}(\textbf{x})-f(\textbf{x}) ||_{2}^{2} \leq C\left( 2^{2J}n^{-\frac{2\beta}{2\beta+p}}+2^{-2J(N+1)}+n^{-\frac{\beta}{2\beta+p}}2^{-J(N+1)}\right)\,,
\end{equation}
from (\ref{eq:ProofProp5.6}) and (\ref{eq:proofProp6.2}).

\medskip

The last result implies that it is possible to choose $J=J(n)$ such that the upper bound of the Risk is minimized. Consequently, (ignoring constants) it is possible to show that $2^{J(n)}\simeq n^{\frac{2\beta}{(2\beta+p)(N+3)}}$ provides such optimal result. Moreover, since the upper bound is valid $\forall f\in \mathcal{F}$:

\medskip

\begin{equation}\label{eq:proofProp7.2}
\mathop{\sup}\limits_{f\in \mathcal{F}}\left(\mathbb{E}\left[||\hat{f}_{J}(\textbf{x})-f(\textbf{x}) ||_{2}^{2}\right] \right)\leq \tilde{C}n^{-\left(\frac{2\beta}{2\beta+p}\right)\left(\frac{N+1}{N+3}\right)}
\end{equation}\,,
\medskip

which completes the proof.

\medskip

Under the optimal choice of $J(n)$, it follows:

\begin{eqnarray}
\label{eq:proofProp7.3}
\mathbb{E}\left[\left\|\hat{f}_{J}(\textbf{x})-\mathbb{E}[\hat{f}_{J}(\textbf{x})]\right\|_{2}^{2}\right] & = & \mathcal{O}\left(n^{-\left(\frac{N+2}{N+3}\right)}n^{-\left(\frac{p}{2\beta+p}\right)}\right)\,, \\
\label{eq:proofProp7.4}
\left\|\mathbb{E}[\hat{f}_{J}(\textbf{x})]-f(\textbf{x}) \right\|_{2}^{2} & = & \mathcal{O}\left(n^{-\left(\frac{2\beta}{2\beta+p}\right)\left(\frac{N+1}{N+3}\right)}\right)\,.
\end{eqnarray}
\medskip

As can be observed in (\ref{eq:proofProp7.3}) and (\ref{eq:proofProp7.4}), the variance term of the estimator $\hat{f}_{J}(\textbf{x})$ is influenced primarily by the properties of the functional space that contains $\left\{f_{l}(x)\,,\,l=1,...,p\right\}$ and the wavelet basis $\left\{ \phi^{per}_{J,k}(x), k =0,...,2^{J}-1\right\}$. Similarly, for $n$ sufficiently large, the bias effect dominates in the risk decomposition and is responsible for the average approximation error of the estimator.

\end{document}